\baselineskip=12pt
\magnification 1200
\vsize=8.5 truein
\hsize=5.8 truein
\hoffset=1.0 cm
\overfullrule = 0pt

\font\ftit=cmbx10
\parskip=6pt
\parindent=2pc 
\font\titulo=cmbx10 scaled\magstep1

% lo siguiente define el cuadrito de teorema probado QED

\def\section#1{\vskip 1.5truepc plus 0.1truepc minus 0.1truepc
	\goodbreak \leftline{\titulo#1} \nobreak \vskip 0.1truepc
	\indent}
\def\frc#1#2{\leavevmode\kern.1em
	\raise.5ex\hbox{\the\scriptfont0 $ #1 $}\kern-.1em
	/\kern-.15em\lower.25ex\hbox{\the\scriptfont0 $ #2 $}}

%The following command places a small circle over a character:

  %"equal by definition"
   %short for "one half"

%The following defines the character for Lie derivative. The quantity
%with respect to which the Lie derivative is taken should follow the
%command "\Lie", and is the one-parameter on which the def. depends:

   %semi-direct product
\def\IR{{\rm I\!R}}  %for real numbers
\def\IN{{\rm I\!N}}  %for integer numbers
 %for complex numbers \def\covD{{\rmI\!D}} 

%for covariant derivative "D" \def\Dirac{D\!\!\!\!/} 
%for Dirac's operator "slashed D"
\newbox\pmbbox
 \def\pmb#1{{\setbox\pmbbox=\hbox{$#1$}%
\copy\pmbbox\kern-\wd\pmbbox\kern.3pt\raise.3pt\copy\pmbbox\kern-\wd\pmbbox
\kern.3pt\box\pmbbox}}

%   Si pones en tus definiciones las siguientes instrucciones:

\font\cmss=cmss10
\font\cmsss=cmss10 at 7pt

\def\IZ{\relax\ifmmode\mathchoice
{\hbox{\cmss Z\kern-.4em Z}}{\hbox{\cmss Z\kern-.4em Z}}
{\lower.9pt\hbox{\cmsss Z\kern-.4em Z}}
{\lower1.2pt\hbox{\cmsss Z\kern-.4em Z}}\else{\cmss Z\kern-.4em Z}\fi}

%   podras usar  $ \IZ $  para denotar los enteros.
%   Si pones en tus definiciones las siguientes instrucciones:
\font\cmss=cmss10
\font\cmsss=cmss10 at 7pt
\def\IS{\relax\ifmmode\mathchoice
{\hbox{\cmss S\kern-.4em S}}{\hbox{\cmss S\kern-.4em S}}
{\lower.9pt\hbox{\cmsss S\kern-.4em S}}
{\lower1.2pt\hbox{\cmsss S\kern-.4em S}}\else{\cmss S\kern-.4em S}\fi}
%   podras usar  $ \IS $  para denotar los numeros de Stirling de
%   segunda clase

% Hay que usar lo que sigue a continuacion para hacer el caracter
% bold     "  \pmb4   "
% ESTAMOS HACIENDO LAS SIGUIENTES DEFINICIONES
% \IN = {0, 1, 2,...}
% \IZ^{+} = {1, 2,...}

% \preprintno{ICN-UNAM-96-10; chao-dyn/9610005}
% \preprintno{September 28, 1996}
% \parindent 0pt

\centerline{\ftit DISCRETE DYNAMICAL SYSTEMS EMBEDDED IN CANTOR SETS}

\vskip 0.5pc

\centerline{Fabio Benatti}

\centerline {Dipartimento di Fisica Teorica}
\centerline{Universit\`a di Trieste}
\centerline{Strada Costiera 11, 34014 Trieste, Italy}
\centerline{and Istituto Nazionale di Fisica Nucleare, Sezione di Trieste}
\centerline{e-mail: fabio.benatti@ts.infn.it}
 
\vskip 0.5pc
 
\centerline{Alberto Verjovsky and Federico Zertuche}

\centerline{Instituto de Matem\'aticas, UNAM}
\centerline{Unidad Cuernavaca, A.P. 273-3}
\centerline{62251 Cuernavaca, Morelos, M\'exico.}
\centerline{e-mail: alberto@matcuer.unam.mx}
\centerline{e-mail: zertuche@matcuer.unam.mx}

\vskip 0.5pc

\vskip 0.5pc

\centerline {Abstract}  
{\leftskip=1.5pc\rightskip=1.5pc\noindent While the notion of chaos is
well established for dynamical systems on manifolds, it is not so for
dynamical systems over discrete spaces with $ N $ variables, as binary
neural networks and cellular automata. The main difficulty is the
choice of a suitable topology to study the limit $ N \to
\infty $. By embedding the discrete phase space into a Cantor set we
provided a natural setting to define topological entropy and Lyapunov
exponents through the concept of error-profile. We made explicit
calculations both numerical and analytic for well known discrete
dynamical models.}

\baselineskip = 12pt

\

\noindent Short Title: {\it Discrete Dynamical Systems}

\


\


\vfill\eject

\baselineskip = 12pt

\section{1. Introduction} 

For motions on differentiable manifolds, the commonly accepted notion
of chaos identifies it with the so-called {\it sensitive dependence on
initial conditions} and the latter with the existence of positive
Lyapunov exponents signaling exponential separation of initially
close trajectories~${}^{1}$.  In this sense, chaotic motion means
unstable behavior; there is, however, an equivalent interpretation of
chaos in terms of information production~${}^{2}$.  This is due to a
celebrated theorem by Pesin which says that, for sufficiently regular
ergodic systems, the sum of the positive Lyapunov exponents coincides
with the {\it Kolmogorov-Sinai dynamical (KS-)entropy} associated with
the dynamics.  The KS-entropy measures the long run unpredictability
of the motion with respect to an invariant state. Further, a
variational principle states that the maximal KS-entropy with respect
to all possible invariant states of a homeomorphism on a compact
metric space is Bowen's {\it topological entropy} which gives a
state-independent description of the degree of chaos based on how open
sets change during the motion.

Roughly speaking, in standard dynamical system contexts, chaos reveals
itself through the exponential increase of errors or, equivalently,
via not-less-than linear information production.

On the contrary, there is no definite agreement about what chaos
should mean in discrete dynamical systems, such as binary neural
networks or cellular automata, where one cannot directly appeal to
differentiability and thus to the standard definition of Lyapunov
exponents.

However, inspired by the equivalent manifestations of chaos, briefly
sketched above, one may try to overcome the lack of differentiable
structures by looking at entropy-like quantities.

In the following, we shall investigate how far chaos in discrete
system can be identified with the exponential increase of initial
errors or with (topological) information production.

Discrete, deterministic dynamical systems, consisting of $ N $ binary
variables, have finite, even though very large ($2^N$), number of
states~${}^{3-9}$.  This means that their dynamics is eventually going
to end up in a periodic cycle.  Due to this fact, there is no room for
chaotic behavior as it is usually intended, unless the number of
states $N\to\infty$.

In numerical studies of continuous systems one needs to discretize the
manifold in order, to solve physical models based on differential
equations~${}^{10}$.  Once the space has been discretized, the number
of available states is finite and one has no longer a chaotic system
since the motion eventually becomes periodic. However the discretized
systems inherit the natural distance of their continuous limit, so
that, if the number of states goes to infinity, one expects to
smoothly retrieve the continuous structure with all its dynamical
properties. The problem with discrete dynamical systems is that the
``natural'' distance, the so-called Hamming distance is ill-defined in
the limit $ N \to \infty $~${}^{9}$.

To overcome these difficulties; in this work we try to resort to
topological techniques. On compact sets, one may define the concept of
sensitive dependence on initial conditions (together with topological
transitivity) by only topological means and no differentiable
structure. So, we define the topology of a Cantor set and endow it
with compatible metrics that remain well-defined in the limit of
infinitely many states~${}^{9}$. Then we look at the various dynamical
patterns that appear and try to characterize them by adapting standard
tools from ergodic theory as, Lyapunov exponents and topological
entropy~${}^{1,11-14}$.

In Sec.~2, we review the concept of sensitive dependence on initial
conditions formulated in a topological way. We endow the space with
the topology of a Cantor set and introduce metrics compatible with
it. In Sec.~3, Lyapunov exponents in discrete systems are defined by
means of metrics and also in terms of the derivative of suitable
embedding homeomorphisms into the reals. In Sec.~4, the topological
entropy is formulated in terms of spanning sets; while in Sec.~5,
Lyapunov exponents and topological entropy are related to an
appropriate indicator of error propagation, that we call {\it
error-profile}. In Sec.~6, some concrete calculations are presented
and in Sec.~7, conclusions are drawn and future directions of
investigation briefly mentioned.
\bigskip

\section{2. Defining Chaos on Discrete Systems}
\medskip

We shall study discrete systems described in the following
way~${}^{3-9}$: a phase space is defined by a set $\Omega$ of states
${\bf S}$ consisting of $ N $ bits $S_i = \left\{ 0, 1 \right\} $; $ i
= 1,\dots , N $, which evolve according to binary functions $ f:\Omega
\rightarrow \Omega $,
        $$ 
	S_i \left( n + 1 \right) = f_i\left( {\bf S} \left( n \right)
	\right) \ , \eqno(1)
	$$ 
that update each bit $S_i(n)$ at each stroke of time $n$.

In neural networks and cellular automata, in general, all bits have an
equally important role in the development of the system with time.
When the number $N$ of bits is finite, the metric most suited to this
state of affairs is the Hamming distance~${}^{4,5,9}$ given by
	$$ 
	d_H \left( {\bf S, S'} \right) = \sum_{i=1}^N \mid S_i - S_i'
	\mid, \eqno(2) 
	$$ 
for any two states $ {\bf S, S'} \in \Omega $. Note that Hamming
distance counts the number of different bits between ${\bf S}$ and
${\bf S}'$, but it is not sensitive to where the differences occur.

The usual way of identifying chaos in the evolution law (1) is to
study the so-called {\it damage spreading}~${}^{5-7,15}$. One
follows the dynamical development of two states with initial Hamming
distance equal to one and studies how it increases with time $ n
$. The speed of damage spreading is then defined by~${}^{15}$:
        $$      
        v(S,S') = \lim_{n \to \infty} {d_H \left( {\bf S}(n), {\bf
	S}'(n) \right) \over n} \ . \eqno(3)
	$$

Two observations are necessary at this point.  The first is that the
above definition does not correspond to the identification of an
exponential increase of an initial small error, but only discriminates
between sub-linear, linear and super-linear increase of the Hamming
distance.
 
The second is that, in the definition, it is implicit that the number
$N$ of binary variables is infinite, otherwise there will be
recurrences and the limit in (3) would automatically vanish.  However,
when $N\to\infty$, the Hamming distance makes no sense since there are
infinitely many states with infinite $d_H$, and then it fails to be a
properly defined distance function.
\medskip

\noindent
{\bf Remark 2.1}\quad We stress that if one wants to think of binary
systems as discretizations of continuous ones, so that asymptotic
quantities like (3) make sense, then the metric of the space should be
well defined when $ N \to \infty $. There are two alternatives: either
binary systems are taken as intrinsically discrete, in such a case
formula (3) is to be investigated as a possible behavior over finite
time-scales~${}^{16}$. Or the number of states is allowed to go to
infinity, in such a case, appropriate metrics that are well defined
for $ N \to \infty $ have to be chosen in order to look at the
dynamics from a topological point of view.

\medskip

In this paper we are going to explore the topological point of
view. Let us take $ N \to \infty $ and introduce the base of open
sets~${}^{17}$
        $$
	{\cal N}({\bf S},q)=\Bigl\{{\bf S}'\in\Omega \mid \
	S_k = S'_k\,,\ 1 \leq k < q\Bigr\}\ .
	\eqno(4)
	$$
It is well known that they generate the topology of a Cantor set on
$\Omega$~${}^{9,11,12}$.

\noindent
{\bf Definition~2.1}: 
A Cantor set is a topological space such that~${}^{11}$:

\noindent
$i)$ it is {\it totally disconnected};

\noindent
$ii)$ {\it perfect}, that is, it is closed and all its points are
accumulation points;

\noindent
$iii)$
{\it compact}.
\smallskip

The main issue in what follows is the identification of chaotic
behaviors in discrete systems with a Cantor structure.  As observed
in the introduction, the lack of differentiability excludes
that one may recognize any exponential separation of trajectories from
the analysis of the tangent map.

Before trying to explore the possible existence of an exponential
increase of initial small errors, one may start from a weaker form of
instability than the usual one and identify a minimal degree of
chaoticity with the following topological definition~${}^{9}$.
\medskip

\noindent 
{\bf Definition~2.2}: Let $f: \Omega \rightarrow \Omega $ be a
continuous map; we say that it shows {\it weak sensitive dependence on
initial conditions (w.s.d.i.c.)} if there exists a $ p \in \IN $ such
that for any $ {\bf S} $ and any $ {\cal N} ({\bf S},q) $, there
exists a ${\bf S}'\in {\cal N} ({\bf S},q) $ and a $ k \in \IN $ such
that \break $ f^k({\bf S}')\notin {\cal N} (f^k({\bf S}), p) $.
\medskip

\noindent
{\bf Remark 2.2} \quad Note that the only requisite to define {\it
w.s.d.i.c} is to have a topology on $\Omega$. Which one? It depends
not on mathematical arguments, but on physical considerations. That
is, which properties do we want to measure and with how much accuracy?
As extreme examples: in the {\it trivial topology} (given by $ \left\{
\emptyset,\Omega\right\} $) the dynamics is going to be trivial;
while, in the {\it discrete topology} (where any subset of $\Omega$ is
an open set) all systems show {\it w.s.d.i.c}.

\medskip

There are several metrics compatible with the topology generated by
the base (4), the more popular is~${}^{11,12}$
	$$ 
	\tilde{d} \left( {\bf S, S'} \right) = \sum_{i=1}^\infty
	{1 \over 2^{i-1}} \mid S_i - S_i' \mid .
	$$ 
However, for the purpose of this work we are going to use, for any $
0 < \beta < 1 $, the following ones
	$$
	d_\beta \left( {\bf S}, {\bf S'} \right) = \beta^m
	\hskip 11pt \hbox{if} \hskip 9pt S_k = S_k' \qquad \forall \ 1
	\leq k < m \hskip 9pt \hbox{and} \hskip 9pt S_m \not= S_m'\ ,
	\eqno(5) 
	$$
in terms of which the base (4) can be expressed as
        $$
	{\cal N}({\bf S},q)=\Bigl\{{\bf S}'\in\Omega \mid \
	d_\beta \left( {\bf S, S'} \right) \leq \beta^q \Bigr\}\ .
	\eqno(6)
	$$
\medskip

\noindent
{\bf Remarks 2.3}
\item{{\it i})}
We can view the embedding process of a finite discrete system into the
Cantor set as follows: Let $ {\cal F} $ be the set of all the
continuous functions $ f: \Omega \to \Omega $ on the Cantor set. Then
any finite discrete dynamical system with $ N $ bits described by
equations (1) is an element of the set
$$
{\cal F}_N = \left\{ f \in {\cal F} \vert \ \forall \ {\bf S} \in
\Omega \ \ f \left( {\cal N} \left( {\bf S}, N \right) \right) = {\cal
N} \left( f \left( {\bf S} \right), N \right) \right\}. 
$$
\item{{\it ii})}
There is a price to pay for working with the
base~(4) and their associated metrics (5). Indeed, some of the binary
variables contribute more than others.  At first sight, this looks as
a major problem since typical binary systems such as the ones
constructed with random couplings~${}^{8,18,19}$ evolve through
functions $f_i$'s where all the variables contribute on an equal
footing to the dynamics and so, apparently, there is no reason to
``dismiss'' some and ``privilege'' others as the distances (5)
do. However, if we make a permutation $ \pi:\IN \to \IN $ of the
automata's indexes, and thus re-enumerate them, the induced mapping $
\widehat{\pi}: \Omega \to \Omega $, such that
$$ 
\widehat{\pi} \left(
S_1, S_2, \dots \right) = \left( S_{\pi^{-1} (1)}, S_{\pi^{-1} (2)},
\dots \right), \eqno(7)
$$ 
is, as we show below, an homeomorphism and so the Cantor topology is
preserved despite the fact that the new metrics are not going to be
Lipschitz equivalent~${}^{13}$. From this it follows that, since {\it
w.s.d.i.c.} is a topological property, it does not depend in which way
we have numerated the automata.

Let us now show that $ \widehat{\pi} $ is an homeomorphism: By
construction the function is a bijection so we only need to show that
it is continuous. Given $ \varepsilon > 0 $ choose $ M \in \IN $ such
that $ \beta^M < \varepsilon $. Now take $ m \in \IN $ such that $
\pi^{-1} \left( i \right) < m $ for any $ 1 \leq i < M $. Then, 
$$
d_\beta \left( {\bf S}, {\bf S}' \right) < \beta^m 
\Longrightarrow d_\beta \left(
\widehat{\pi} \left( {\bf S} \right), \widehat{\pi} \left( {\bf S}' \right)
\right) \leq\beta^M < \varepsilon,
$$
hence continuity.
\medskip

\noindent
{\bf Remarks 2.4}

\item{{\it i})} The metrics (5) seem to establish a preferred direction 
along the network. However, one can always reverse it by means of the
permutation
$$
\widehat{\pi} \left( S_1, S_2, \dots, S_N \right) = \left( S_N,
S_{N-1}, \dots, S_2, S_1 \right) 
$$
and, then, take the limit $ N \to \infty $ in the metrics (5).

\item{{\it ii})} The metrics (5) are suited to semi-infinite
networks. However, when expanding to infinity finite networks with
periodic boundary conditions, as we will see in the examples, more
symmetric metrics are preferable. These are achieved by means of
two-sided sequences $ {\bf S} = \left( \dots S_{-2}, S_{-1}, S_0, S_1,
S_2, \dots \right) $ and using the metrics
	$$
	\eqalign{\widehat{d}_\beta \left( {\bf S}, {\bf S'} \right) =
	\beta^m \hskip 11pt & \hbox{if} \hskip 9pt S_k = S_k' \qquad
	\forall |k| < m \cr & \hbox{with} \hskip 9pt S_{m} \not=
	S_{m}' \hskip 9pt \hbox{or} \hskip 9pt S_{-m} \not= S_{-m}'\
	\cr} 
	$$
which also define a Cantor topology.
\medskip

\section{3. Lyapunov Exponents}

Since Definition 2.2 is a topological one, and the metrics (5) all
define the same topology, we can use any of them (by fixing a $ \beta
$) to check if there is {\it w.s.d.i.c.} or not.

However, in continuous dynamics there is a definition of {\it
sensitive dependence on initial conditions}, which we shall refer to
as {\it strong (s.s.d.i.c)} in comparison with the previous one ({\it
w.s.d.i.c.}), which is based on the concept of positive Lyapunov
exponents; that is, on the exponential separation of initially close
trajectories~$^{1,11,13}$. Such a behavior is usually associated
with exponential increase of initial small errors. We propose two
different natural definitions of Lyapunov exponents. The first one is
based on the metrics (5), the second one is via an embedding of $
\Omega $ into the continuum; where the notion of derivative can be
used. 

The metrics (5) offer a natural means to measure the increase of
errors in Cantor sets.  In fact, given $d_\beta$, one can define
Lyapunov exponents as follows:
	$$ 
        \lambda_M({\bf S}) = \limsup_{n \to \infty} \lim_{d_\beta({\bf
	S},{\bf S'})\to 0} \ {1 \over n} \log_{\beta^{-1}}
	{d_\beta\Bigl( f^n({\bf S}),f^n({\bf S}')\Bigr) \over
	d_\beta({\bf S},\bf{S}')}\ .  \eqno(8)
        $$ 
The quantity $ \lambda_M({\bf S}) $ depends in general on ${\bf S}$:
it amounts to identify separation of trajectories with the following
behavior
$$
d_\beta\Bigl(f^n({\bf S}),f^n({\bf S}')\Bigr)\simeq
\beta^{-n\,\lambda_M({\bf S})} d_\beta({\bf S},{\bf S}')\ . 
$$

\noindent
{\bf Remarks 3.1}

\item{{\it i})} In (8), we have used $\limsup$ as we do not know whether the
limit for $n\to\infty$ exist as it is the case for smooth dynamical
systems by Oseledec's multiplicative theorem~${}^{2}$. Also, the limit
when $ d_\beta({\bf S},{\bf S'})\to 0$ may very well diverge as is the
case when discrete dynamical systems exhibit nearly stochastic
behavior such as random boolean networks or binary neural networks
with long range connections among the variables~${}^{8,9,15,18}$.

\item{{\it ii})} Since the distances depend on $ \beta $, we use a
logarithm base $ \beta^{-1} $ to make $ \lambda_M ({\bf S}) $ $ \beta
$ independent.

\item{{\it iii})} Due to the presence of a positive $\lambda_M $
because of the exponential separation of trajectories, however close
to each other initially, it turns out that the {\it s.s.d.i.c.}
property implies the weaker {\it w.s.d.i.c.} property.
\medskip

One may also try a kind of differential approach to the notion of
exponential instability which is based on an appropriate embedding of
the Cantor set $\Omega$ into the reals (compare, the abstract
mathematical approach in ${}^{20}$). Let us consider the following
commutative diagram
	$$ \eqalign{&\Omega \buildrel f \over \longrightarrow \Omega
	\cr {}_\phi & \downarrow \ \ \quad \downarrow {}_\phi \cr &
	\Xi \buildrel \tilde{f} \over \longrightarrow \Xi \cr}
	\eqno(9) 
	$$ 
which defines the function $ \tilde{f} $, by means of a homeomorphism
$ \phi $, with $ \Xi $ being a Cantor set embedded in the reals ($ \Xi
\subset \IR $). It is important to observe that due to the
commutativity of the diagram (9), the dynamics generated by $ f $ and
$ \tilde{f} $ are intrinsically the same because of {\it topological
conjugacy}~${}^{11,12}$. Since $ \Omega $ is an uncountable compact
Abelian topological group; there are uncountably many ways of
constructing a homeomorphism $ \phi: \Omega \to \Xi $ (for instance,
by suitable translations)~${}^{21}$. We will consider a $\phi$ that is
suited to the metrics (5). Explicitly, let
	$$ \phi \left( {\bf S}\right) = \sum_{k = 1}^\infty \gamma_k
	S_k\ , \eqno(10a) $$
where 
        $$
	\gamma_k = h^k \left( h^{-1} - 1 \right) =
	\Bigl({1-\alpha\over 2}\Bigr)^k{1+\alpha\over 
	1-\alpha} \ , \eqno(10b) 
	$$

\

\noindent
with
        $$
	h = {1-\alpha\over 2} \hskip 11pt \hbox{and} \hskip 9pt 0 <
	\alpha < 1 \ ,
	$$
sets the scale of the Cantor set by suppressing intervals in the
proportion $ \alpha $ (the standard choice being $ \alpha = 1/3 $).
Figure~1 explains the idea of the construction of the Cantor
set in a graphical way. Note that the self-similar nature of the
Cantor set is reflected by the fact that the coefficients $
\gamma_k $ satisfy the following recursion relation
	$$    
        \gamma_{k+m} = h^m \ \gamma_k \ . \eqno(11) 
        $$
Now, given any continuous function $ f:
\Omega \to \Omega $, we define the function \break $ \delta_h f:\Omega \to
\IR $ by
        $$ 
	\delta_h f\left( {\bf S} \right) = \lim_{\bf S' \to S} {\phi
	\circ f \left( {\bf S'} \right) - \phi \circ f \left( {\bf S}
	\right) \over \phi \left( {\bf S'} \right) - \phi \left( {\bf
	S} \right)}, \eqno(12)
	$$	
which has the typical properties of a derivative. In particular it
maps to the vector space $\IR $. Indeed, (12) is nothing but the
derivative of the conjugate map $ \tilde{f} $.  Of course the actual
value of $ \delta_h f({\bf S})$ is $\alpha$-dependent, for if one
wants to give the instantaneous rate of change of a function, one
needs a scale!  Similarly, if one wants to speak about the Hausdorff
dimension of a Cantor set, one needs to embed it into $ \IR $ and
the result is going to be scale-dependent.

From (12), there naturally comes the following proposal of
Lyapunov exponent associated with the derivative: 
	$$ 
        \lambda_D \left( {\bf S} \right) = \limsup_{n \to \infty} {1
	\over n} \log_{h^{-1}} \left\vert \delta_h f^n \left( {\bf S}
	\right) \right\vert \ . \eqno(13)
        $$ 

\section{4. Entropy}
\medskip

In ergodic theory, one approaches the notion of entropy from two
different perspectives: the first one is statistical and based on the
presence of an invariant measure, the other is topological. We shall
consider the latter point of view which leads to the notion of
topological entropy~$^{1,13,14}$.
\bigskip

\section{4.1 Topological Entropy}
\medskip

In the topological case, the fundamental objects are the open sets
(4). We shall calculate $h_{top}(f)$ following standard
techniques~$^{1,13}$, namely the so-called $(n,\varepsilon)-${\it
spanning set}.  For this we need the dynamics-dependent distances
$$
d_{\beta,n}({\bf S},{\bf S}')=
\max_{0\leq k\leq n}d_\beta\Bigl(f^k({\bf S}),f^k({\bf S}')\Bigr)\ ,
$$
and the corresponding open balls
        $$
	{\cal B}_\beta({\bf S},\varepsilon,n)=\Bigl\{{\bf S}'\in\Omega \mid 
	d_{\beta,n}({\bf S},{\bf S}')<\varepsilon\Bigr\}\ .
	\eqno(14)
	$$
A subset $ E \left( n, \varepsilon \right) \subseteq\Omega$ is called
$ (n,\varepsilon) $-spanning if
        $$
	\Omega = \bigcup_{{\bf S}\in E \left( n, \varepsilon
	\right)}{\cal B}_\beta({\bf S},\varepsilon,n)\ .
	\eqno(15)
	$$
That is, any $ (n,\varepsilon)$-spanning set corresponds to an open
cover of $\Omega$.  Because of the compactness of $\Omega$, there will
always be an $(n,\varepsilon)-$spanning set containing finitely many
states ${\bf S}$; therefore, the minimal cardinality
        $$
	{\cal S}(\varepsilon, n)=\min_{E \left( n, \varepsilon \right)}
	\hskip 0.2cm \# \ E \left( n, \varepsilon \right) 
	$$
of $(n,\varepsilon)-$spanning sets is finite.

The topological entropy $h_{top}(f)$ is defined by
$$
h_{top}(f)=\lim_{\varepsilon\to 0}\limsup_{n\to\infty}{1\over n}
\log{\cal S}(\varepsilon, n)\ ,
\eqno(16)
$$
where the logarithm is in base $2$.

Since the topological entropy $h_{top}(f)$ reflects the way open sets
change in time under $f:\Omega\to\Omega$ given by (1), $h_{top}(f)$
will not depend on $\beta$. This is Bowen's definition and is based on
using metrics~${}^{13}$. The intrinsically topological nature of the
notion rests on the fact that Bowen's formulation is equivalent to the
one of Adler, Konheim and Mc Andrew based on open covers~${}^{1,13,14}$.
\medskip

\noindent
{\bf Remark 4.1}

The topological entropy $h_{top}(f)$ is related by a variational
principle to the metric or dynamical entropy of Kolmogorov,
$h_\mu(f)$:
$$
h_{top}(f)=\sup_{\mu\in{\cal M}(\Omega,f)} h_\mu(f)\ ,
\eqno(17)
$$
where $ {\cal M} (\Omega,f) $ is the space of invariant measures in $
\Omega $ under the dynamics  $ f $. To establish (17) one needs a
measurable structure to be defined on $\Omega$, which is easily
achieved by considering the $\sigma$-algebra generated by the open
sets (4). What is more difficult to obtain is a measure $\mu$ on the
$\sigma$-algebra being invariant under the dynamics:
$\mu(f^{-1}(C))=\mu(C)$ for all measurable subsets $C\subseteq\Omega$.
This is the meaning of the request $ \mu\in{\cal M}(\Omega,f) $. Such
a problem will be matter of further investigation and will not be of
concern in this work.

\section{5. Error-profile and chaotic behavior}

Now we are going to study how Lyapunov exponents (8), (13) and the
topological entropy (16); are related to error propagation along the
ordering defined by the metrics (5) over the network.

Let us take two near states $ {\bf S} $ and $ {\bf S'} $ with initial
distance 
         $$ 
	 d_\beta \left( {\bf S, S'} \right) = \beta^q \ .  \eqno(18a) 
	 $$
Their evolution in time can always be written as
	$$ 
	d_\beta \Bigl( f^n({\bf S}), f^n({\bf S'})\Bigr) 
	= \beta^{q - L_n \left( {\bf S, S'}
	\right)}\ , \eqno(18b) 
	$$ 
where $ L_n \left( {\bf S, S'} \right)\in \IZ $ measures the length
traveled to the left ($L_n>0)$, or to the right ($L_n<0$), by the
errors at the $n$--th time step. The behavior of $ L_n \left( {\bf S,
S'}\right) $, numerically measurable, reflects the properties of the
network dynamics and is not necessarily monotonically increasing with
time.
\smallskip

\noindent
{\bf Remark 5.1}\quad It is to be emphasized that $ L_n \left( {\bf S,
S'} \right) $ does not correspond to the Hamming distance, since $ q -
L_n \left( {\bf S, S'} \right) $ locates the first error that appears
in the automaton ordering associated with the metrics (5).  Such an
error may very well be the only one, in such a case: $d_H \left(
f^n({\bf S}), f^n({\bf S'}) \right) = 1$.
\smallskip

Thus, the picture we have in mind is as follows.
Let us assume $L_n>0$:  at time $n=0$, take
two states ${\bf S}$, ${\bf S'}$ which agree upon the first $q - 1$
bits, at time $n=1$ they agree upon the first $q - L_1\left({\bf
S},{\bf S}'\right) -1 $ bits, at time $n=2$ upon the first $q - L_2
\left( {\bf S, S'} \right)-1 $ bits, and so on. In this way, after $n$
iterations of the dynamics (1), the first error will have propagated from
position $ q $ to position $ q - 1-L_n \left( {\bf S, S'} \right) $.

\noindent
{\bf Remark 5.2} \quad If $L_n<0$, then the initial error moves
further and further away to the right with two consequences: first, it
need not be the first error and thus need not appear at the exponent
in (18b); second, it may become smaller and smaller contrary to the
expectation that instability should amplify initial small errors.
However, this is due to a preferred direction inherent in the choice
of the metrics (5) as discussed in Remark 2.4.{\it i}. This also means
that we can always consider $ L_n > 0 $ in (18b) up to a reflection:
this argument particularly applies to the behavior of the rule 30 in
Wolfram's classification (see the examples) which seems otherwise to
contradict instability.
\medskip

\noindent
{\bf Definition 5.1}\quad We define the ${\bf S}$-{\it error-profile}
by the following limit,
	$$
	\Lambda_n \left( {\bf S} \right) = 
        \limsup_{d_\beta\left({\bf S},{\bf S}'\right) \to 0} 
	L_n\left( {\bf S, S'} \right)\ , \eqno(19) 
	$$  
measures the length traveled by the errors at the 
$n$--th time step due to two infinitesimally closed initial states. 
\medskip

\noindent
{\bf Remark 5.3}\quad
The idea behind the previous definition is that, in physical 
instances, $L_n\left( {\bf S}, {\bf S}'\right)
\approx \Lambda_n \left({\bf S}\right)$ once spurious boundary effects
are eliminated by ${\bf S}' \to {\bf S}$ in the Cantor topology
defined by the metrics (5). Numerically, the limit in (19), will be
later handled by considering ${\bf S}$ and ${\bf S}'$ with
$d_\beta({\bf S},{\bf S'})$ sufficiently small.
\medskip

We are now going to see how the error-profile is related with the
concepts introduced in the previous sections. First, we deal with the
Lyapunov exponents introduced in Sec.~3.

\section{5.1 Lyapunov Exponents}
\medskip

Concerning metric-based Lyapunov exponents $\lambda_M$  defined 
in~(8), equations (18) and (19) yield 
	$$ 
        \lambda_M \left({\bf S} \right) = \limsup_{n \to \infty}
	{\Lambda_n \left( {\bf S} \right) \over n}\ . \eqno(20) 
	$$ 

Concerning derivative-based Lyapunov exponents $\lambda_D$ defined
in~(13), we first calculate the derivative of a generic continuous
function $ f:\Omega \to \Omega $.

Consider two closed points $ {\bf S, S'} \in \Omega $ with distance $
\beta^m $, they are of the form (see equation (5)):
	$$ {\bf S} = \left( S_1 S_2 S_3 \dots S_{m-1} S_m S_{m+1}
	\dots \right) \eqno(21a)$$
and 
	$$ {\bf S'} = \left( S_1 S_2 S_3 \dots S_{m-1} S'_m S'_{m+1}
	\dots \right) \ , \eqno(21b) 
	$$ 
where $ S_m \not= S'_m $. Applying the homeomorphism (10a), we get
	$$ 
	\Delta {\bf S} \equiv \phi \left( {\bf S'} \right) - \phi
	\left( {\bf S} \right) = \sum_{k=m}^\infty \gamma_k \ \left(
	S_k' - S_k \right). \eqno(21c)
	$$ 
The images of $ {\bf S} $ and $ {\bf S}' $ after the $n$-th time-step
$ f^n: \Omega \mapsto \Omega $ are; in a short hand notation,
	$$ f^n \left( {\bf S} \right) = \left( \zeta_1 \zeta_2 \zeta_3
	\dots \zeta_{p_n-1} \zeta_{p_n} \zeta_{p_n+1} \dots \right)
	$$ 
and
	$$ 
	f^n \left( {\bf S}' \right) = \left( \zeta_1 \zeta_2 \zeta_3
	\dots \zeta_{p_n-1} \zeta_{p_n}' \zeta_{p_n+1}' \dots 
	\right)\ ,  
	$$ 
where $ \zeta_{p_n} \not= \zeta'_{p_n} $, with $ p_n = p_n \left( m
\right) $ being a function, of the initial error position $ m $ in
(21). From (21c), we have that 
	$$ \Delta {\bf S} = \gamma_m \ \mu_m + R_m 
	$$
where $ \mu_m \equiv S'_m - S_m = \pm 1 $ and $ R_m \equiv
\sum_{k=m+1}^\infty \gamma_k \ \mu_k$.  
Now, by means of (10b) and (11) follows that $ \mid R_m \mid \leq
\sum_{k=m+1}^\infty \gamma_k = h^m$. So that $ R_m \sim {\cal O}
\left( h^m \right) $ which gives
	$$ 
	\Delta {\bf S} = h^m \left( h^{-1} - 1 \right) \ \mu_m
	\left( 1 + {\cal O} \left( h \right) \right). 
	$$ 
An analogous expression is obtained for $ \Delta f^n \equiv f^n \left(
{\bf S} \right) - f^n \left( {\bf S}' \right) $ giving 
	$$ 
	\Delta f^n = h^{p_n(m)} \left( h^{-1} - 1 \right) \
	\mu_{p_n(m)} \left( 1 + {\cal O} \left( h \right) \right) \ .
	$$
Therefore, 
	$$ 
	\delta_h f^n \left( {\bf S} \right) = \lim_{m\to \infty}
	{\Delta f^n \over \Delta {\bf S}} = \lim_{m\to \infty} \mu_m \
	\mu_{p_n(m)} h^{p_n(m)-m} \left( 1 + {\cal O} \left( h \right)
	\right).
	$$
If the limit exists; we obtain 
	$$ 
	\delta_h f^n \left( {\bf S} \right) = \pm h^{- \Lambda_n({\bf
        S})} \left( 1 + {\cal O} \left( h \right) \right), 
	$$  
where
	$$ 
	\Lambda_n({\bf S}) \equiv
        \lim_{m \to \infty} (m-p_n(m)) \ ,
	$$ 
which, if exists; it is the error propagation at time $n$ as seen
through the embedding (9).

Inserting the above result into (13) yields 
	$$ 
	\lambda_D \left( {\bf S} \right) = \lim_{n \to \infty}
	{\Lambda_n({\bf S}) \over n}\ , \eqno(22) 
	$$
which, if it exists, equals (20).

\noindent
{\bf Remark 5.4}\quad As observed in Remark 3.1.{\it ii}, the metric
and derivative definitions of Lyapunov exponents do not depend on the
specific scale used. The velocity of leftward propagation of errors
should be a dynamical effect independent of the scale, which is
exactly what turns out from above, whence the coincidence of (20) and
(22).

\section{5.2 Topological Entropy}

Consider equations (18) and let us define 
        $$
	L^*_n({\bf S},{\bf S}')\equiv\max_{0\leq k\leq n}L_k({\bf
	S},{\bf S}') \ . \eqno(23)
	$$
From (14) it follows that 
        $$
	{\cal B}_\beta({\bf S},\varepsilon,n)=\Bigl\{{\bf S}'\in\Omega
	\mid \beta^{q - L_n^* \left( {\bf S, S'} \right)} < \varepsilon
	\Bigr\}\ .  
	$$
In the particular case where the dynamics (1) is such that $
L^*_n $ is independent of $ {\bf S} $ and $ {\bf S'} $ let us define 
        $$
	\theta_n \equiv L_n^* \left( {\bf S},{\bf S}' \right) 
	\eqno(24)
	$$
and take $ \varepsilon = \beta^p $. From (6) and (14) it follows
that 
        $$
	\eqalign{{\cal B}_\beta({\bf S}, \beta^p, n) &= \Bigl\{ {\bf
	S}'\in\Omega\, \mid \,\beta^q < \beta^{p + \theta_n
	}\Bigr\}\cr &= \Bigl\{ {\bf S}'\in\Omega\, \mid \, d_\beta
	\left( {\bf S, S'} \right) \leq \beta^{p + \theta_n + 1}
	\Bigr\} \cr &={\cal N} \left( {\bf S}, r_{n, p} \right) ,}
	$$
where 
        $$
	r_{n, p} = p + \theta_n + 1 \ .
	$$
Since $ {\cal N} \left( {\bf S}, r_{n, p} \right) $ are elements of
the base (4), it follows that 
        $$ 
	E \left( n, \beta^p \right) = \Bigl\{{\bf S} \in \Omega \mid
	S_k = 0 \ \forall \ k > r_{n, p} \Bigl\} 
	$$
is an $(n,\beta^p)-$spanning set with cardinality $ 2^{r_{n. p}} $;
which, by construction, is minimal. From (16) we obtain
       $$
       h_{top} (f) = \lim_{p\to\infty}\limsup_{n\to\infty} {r_{n,
       p} \over n} \ .
       $$
So that,
       $$
       h_{top} (f) = \limsup_{n\to\infty}{\theta_n \over n}\ .
       \eqno(25)
       $$

\medskip

In the general case (24) is not valid and the behavior of $ L^*_n
\left( {\bf S, S'} \right) $ depends on the fine details of the
dynamics (1). In such cases, it is still possible to set an upper
bound to the value of the topological entropy. Let us define
        $$
	\eta(n)\equiv\max_{{\bf S},{\bf S}'}L_n^*({\bf S},{\bf S}')\ ,
	$$
with $ {\bf S} $ and $ {\bf S'} $ subject to the constraint
(18a). Then (using (6) and (14)), 
        $$
	\eqalign{{\cal B}_\beta({\bf S}, \beta^p, n) &= \Bigl\{{\bf
	S}'\in\Omega \mid \beta^{q - L_n^* \left( {\bf S, S'} \right)}
	< \beta^p \Bigr\} \cr &\supseteq \Bigl\{ {\bf S}'\in\Omega\,
	\mid \,\beta^{q - \eta(n)} < \beta^p \Bigr\}\cr &\supseteq
	\Bigl\{ {\bf S}'\in\Omega\, \mid \,d_\beta \left( {\bf S, S'}
	\right) \leq \beta^{p + \eta(n) + 1} \Bigr\}\cr &\supseteq
	{\cal N} \left( {\bf S}, t_{n,p} \right) \ , } 
	$$
where 
        $$
	t_{n, p} = p + \eta(n) + 1 \ .
	$$
Now 
        $$ 
	E \left( n, \beta^p \right) = \Bigl\{{\bf S} \in \Omega \mid
	S_k = 0 \ \forall \ k > t_{n, p} \Bigl\} \ ,
	$$
is again an $ (n,\beta^p) $-spanning set with cardinality $
2^{t_{n. p}} $, but we cannot assure that it is minimal. So, from
(16) we obtain
        $$
	h_{top} (f) \leq\limsup_{n\to\infty}{\eta(n)\over n}\ ,
	$$
which assures that, if $\eta(n)$ does not increase as $ n $ or faster,
the topological entropy vanishes.

\section{6. Examples}

\section{6.1 The shift map}

We begin applying the ideas developed so far to the $ v $-shift map $
\sigma_v $ defined by
	$$ 
        \sigma_v \left( S_1 S_2 S_3 \dots \right) = \left( S_{1+v}
	S_{2+v} S_{3+v} \dots \right).  
        $$ 
Consider two points $ {\bf S, S'} \in \Omega $ with distance $ d_\beta
\left( {\bf S, S'} \right) = \beta^q $. Applying $ \sigma_v^n $ we
obtain 
	$$ 
        d_\beta \left( {\bf S}\left( n \right) , {\bf S'} \left( n
	\right) \right) = \beta^{q - v \, n}\ . 
        $$ 
According to (18), (23) and (24), we see that $ \theta_n = v \, n
$; thus, from (25) we have
	$$ 
        h_{top} (f) = v > 0 \ .
        $$
It is also evident that the topological entropy coincides with the
Lyapunov exponents $\lambda_M=\lambda_D$ (see equations (20) and
(22)). 

\section{6.2 Networks with three Inputs}

We study now the evolution rule (1) in the case of interactions
involving three nearest neighbors and impose periodic boundary
conditions; specifically
	$$ 
        S_i \left( n+ 1 \right) = f \left( S_{i-1}\left( n \right),
	S_{i}\left( n \right), S_{i+1}\left( n \right) \right) \hskip
	11pt \hbox{for} 
	\hskip 9pt i = 2, 3, \dots, N-1\ , \eqno(26a)
        $$
	$$ 
        S_1 \left( n+ 1 \right) = f \left( S_N\left( n \right),
	S_1\left( n \right), S_2\left( n \right) \right)\ ,
	\eqno(26b) 
        $$
and
	$$ 
        S_N \left( n+ 1 \right) = f \left( S_{N-1}\left( n \right),
	S_N\left( n \right), S_1 \left( n \right) \right)\ ,
	\eqno(26c) 
        $$
where the transfer function $f$ is the same for all the bits. 

Due to the periodic boundary conditions, the infinite limit of these
network requires the symmetric metrics discussed in Remark 2.4.{\it
ii}: this context accommodates errors propagating both to the left and
to the right.

We are going to study the boolean rules numbered $ 30 $, $ 73$, $ 90 $
and $ 167 $ according to Wolfram's scheme~${}^{5-7}$ which we
explicitly list in table~1. The first three columns give the values of
three adjacent bits and the remaining columns show the corresponding
bits for the two rules. We stress that the rule $ 90 $ is the XOR
rule in the two adjacent bits and it is well known as ``chaotic'' in
Wolfram's terminology.

\

We have consider an automaton consisting of $ N = 1000 $ bits and
start with a random initial state. After a transient of length $ N^2 $
we let the dynamics reach a state $ {\bf S} $. Then we choose a state
$ {\bf S'} $ which differs from $ {\bf S} $ in the $499$, $500$ and $
501$-th bits, and start to measure the speed of error-profile $
{\Lambda_n \over n} $ and the speed of damage spreading $ {d_H \left(
{\bf S(n)}, {\bf S'(n)} \right) \over n} $.  
The main results are plotted in
figures~2-4. 

Figures~2 show the spread of errors as the states $ {\bf
S} $, $ {\bf S'} $ evolve in time, each cross corresponding to a
different bit in the two configurations. 

Figures~3 show the speed of error-profile as a function of time.
According to (20) and (22), they exhibit, for $ n >> 1 $ a Lyapunov
exponent $ \lambda = 1 $ for all the rules but the rule $ 73 $ which
shows $ \lambda = 0 $. Same conclusions can be extracted from the
topological entropy.

Figures~4 show the evolution in time of the speed of damage spreading.
One can see that no clear behavior emerges for $ n >> 1 $ that may
help evaluate the damage spreading according to (3); moreover, even
when, according to the Lyapunov and entropic analysis, the behavior
is complex as is the case with rule $ 90 $, there is instead a clear
tendency of the damage spreading to go to zero.

It is important to observe from figure~2.b that rule $ 73 $ shows a
complex behavior. However it is localized in the sense that it does
not grow with $ N $, so for $ N \to \infty $, $ {\bf S} $ and $ {\bf
S'} $ are on a periodic attractor and so the dynamics is not
chaotic. In contrast the other rules, which are chaotic, spread the
errors along all the bits.

\bigskip

\section{7. Conclusions}

We have endowed the phase space of binary variables with the topology
of the Cantor set in the limit when the number of variables $ N $ goes
to infinity. This embedding of the phase space permits us to
understand the dynamical behavior of binary dynamical systems, much on
the same footing as the ones over differentiable manifolds providing a
mathematically solid framework for discrete systems. One of the
advantages of this approach, is the fact that the distance function
(5) is well defined for finite or infinite $ N $. Despite being the
Hamming distance (2), the most natural distance function over the
space of binary variables $ \Omega $, it has the disadvantage of being
divergent as $ N \to \infty $ on states differing on an infinite
number of binary variables.

We have formalized the notion of Lyapunov exponents for discrete
systems in two related ways: by resorting to metrics compatible with
the Cantor topology and by suitably embedding the Cantor structure into a 
differentiable one.

Guided by the connections between Lyapunov exponents and topological
entropy in continuous system, we have also computed the topological
entropy and compared it with the Lyapunov exponents calculated
according to the given prescriptions.  This has been done in Sec.~5
where we related both notions to the concept of error-profile which is
a phenomenological quantity that can be accessed numerically and has a
self-evident physical interpretation.  We have illustrated all these
concepts by examples in Sec.~6.

Further points that deserve to be studied are:

\noindent
{\it i}) The problem of the concept of a derivative. Here we have
introduced it with the aid of the homeomorphism (10) which is
compatible with the metrics (5). However, from the mathematical
point of view it would be better if one could construct a meaningful
``discrete derivative'' which is homeomorphism free.

\noindent
{\it ii}) The construction of an invariant measure for the definition
of the metric entropy (17), as sketched in Remark~4.1.

\noindent
{\it iii)} The application of the methods presented above to the
treatment of Kauffman's models of cellular automata with connectivity
$K$ and random couplings which show a transition from an ordered phase
for $ K \leq 2 $ where the length of the attractors grows as $ {\sqrt
N} $, to a disordered one, termed chaotic, for $ K > 2 $ with lengths
growing as $ {\rm e}^N $~${}^{3,5,18}$.

\section{Acknowledgments}

The authors thank the referee for his/her accurate criticisms which
contributed to make a better article. Third author (F.Z.) thanks
T. Figueras for useful suggestions while reading the man\-u\-script.
This work is supported in part by {\bf CONACyT} project number
U40004-F.

\vfill\eject

\centerline{{\ftit References}}

\noindent 
\item{${}^{1}$} R. Ma\~n\'e. {\it Ergodic Theory and Differentiable
Dynamics}. Springer, Berlin (1987).

\noindent 
\item{${}^{2}$} J.-P. Eckmann and D. Ruelle {\it 
Ergodic Theory of Chaos and Strange Attractors}. Rev.Mod. Phys. {\bf 57}
(1985) 617.

\noindent 
\item{${}^{3}$} M. Aldana, S. Coppersmith and L. Kadanoff. {\it 
Boolean Dynamics with Random Couplings}. In: Perspectives and Problems
in Nonlinear Science, 23--89. Springer Verlag, New York (2003).

\noindent 
\item{${}^{4}$} J. Hertz, A. Krogh and R.G. Palmer. {\it
Introduction to the Theory of Neural Computation}. Addison-Wesley
Redwood City, CA (1991).

\noindent 
\item{${}^{5}$} G. Weisbuch. {\it Complex Systems Dynamics}. Addison
Wesley, Redwood City, CA (1991). 

\noindent 
\item{${}^{6}$} S. Wolfram. {\it Universality and Complexity in Cellular
Automata}. Physica~D {\bf 10} (1984) 1. 

\noindent 
\item{${}^{7}$} S. Wolfram. {\it Theory and Applications of Cellular
Automata}. World Scientific (1986). 

\noindent 
\item{${}^{8}$} A. Crisanti, M. Falcioni and A. Vulpiani. {\it
Transition from Regular to Complex Behavior in a Discrete
Deterministic Asymmetric Neural Network Model}. J. Phys. A:
Math. Gen. {\bf 26} (1993) 3441.

\noindent 
\item{${}^{9}$} H. Waelbroeck and F. Zertuche. {\it Discrete Chaos}. 
J.~Phys.~A:~Math.~Gen. {\bf 32} (1999) 175-189.

\noindent 
\item{${}^{10}$} K.E. Atkinson. {\it An Introduction to Numerical
Analysis}. John Wiley \& Sons, Singapore (1989).

\noindent 
\item{${}^{11}$} S. Wiggins. {\it Dynamical Systems and Chaos}. Springer,
New York (1990). 

\noindent 
\item{${}^{12}$} R. Devaney. {\it An Introduction to Chaotic
Dynamical Systems}. Addison Wesley Publ. Co. Reading MA, (1989).

\noindent 
\item{${}^{13}$} A. Katok and B. Hasselblatt. {\it Introduction to the Modern
Theory of Dynamical Systems}. Cambridge University Press, Cambridge UK
(1995).

\noindent 
\item{${}^{14}$} P. Walters. {\it Introduction to Ergodic Theory}. Springer
Verlag, New York (1982). 

\noindent 
\item{${}^{15}$} G. Boffetta, M. Cencini, M. Falcioni and A. Vulpiani. {\it
Predictability: a Way to Characterize Complexity}. Phys. Rep. {\bf
356} (2002) 367-474. 

\noindent 
\item{${}^{16}$} F. Benatti, V. Cappellini and F. Zertuche. {\it Quantum
Dynamical Entropies in Discrete Classical Chaos}.
J.~Phys.~A:~Math.~Gen. {\bf 37} (2004) 105-130.

\noindent
\item{${}^{17}$} Y. Choquet-Bruhat, C. DeWitt-Morette and
M. Dillard-Bleick. {\it Analysis, Manifolds and Physics}. Elsevier,
The Netherlands (1982).

\noindent 
\item{${}^{18}$} S.A. Kauffman. {\it The Origins of Order. Self-Organization
and Selection in Evolution}. Oxford University Press (1993).

\noindent 
\item{${}^{19}$} F. Zertuche, R. L\'opez-Pe\~na and H. Waelbroeck. {\it
Storage Capacity of a Neural Network with State-Dependent Synapses}.
J. Phys. A: Math. Gen. {\bf27} (1994) 1575; {\it Recognition of
Temporal Sequences of Patterns with State-Dependent
Synapses}. J. Phys. A: Math. Gen. {\bf 27} (1994) 5879-5887.

\noindent
\item{${}^{20}$} D. Sullivan. {\it Differentiable Structures on Fractal Like
Sets, Determined by Intrinsic Scaling Functions on Dual Cantor Sets},
in: {\it Nonlinear Evolution and Chaotic Phenomena}. 101-110,
G. Gallavotti and P.F. Zweifel eds, Nato Adv. Sci. Inst. Ser.B Phys.,
176, Plenum, New York (1988).

\noindent 
\item{${}^{21}$} L. Pontrjagin. {\it Topological Groups}. Princeton
University Press, Princeton, N.J. (1958).

\eject \vfill

\noindent
Table~1

$$ \vbox{\tabskip= 0pt \offinterlineskip
\halign to300pt{\strut#& \vrule#\tabskip=1 em plus2em& \hfil#&
\vrule#& \hfil#\hfil& \vrule#& \hfil#\hfil& \vrule#& \hfil#\hfil&
\vrule#& \hfil#\hfil& \vrule#& \hfil#\hfil& \vrule#& \hfil#&
\vrule#\tabskip=0pt\cr\noalign{\hrule} 
& & \omit\hidewidth $ S_{i-1} $ \hidewidth& & \omit\hidewidth $S_{i}
$ \hidewidth& & \omit\hidewidth $ S_{i+1} $ \hidewidth& &
\omit\hidewidth $ 30 $ \hidewidth& & \omit\hidewidth $ 73 $
\hidewidth& & \omit\hidewidth $
90 $ \hidewidth& & \omit\hidewidth $ 167 $ \hidewidth&
\cr\noalign{\hrule}& & \ 0 & & \ 0 & & \ 0 & & \ 0 & & \ 1
& & \ 0 & & \ 1 & \cr \noalign{\hrule} & & \ 0 & & \ 0
& & \ 1 & & \ 1 & & \ 0 & & \ 1 & & \ 1 & \cr 
\noalign{\hrule} & & \ 0 & & \ 1 & & \
0 & & \ 1 & & \ 0 & & \ 0 & & \ 1 & \cr
\noalign{\hrule} & & \ 0 & & \ 1 & & \ 1 & & \ 1 & & \ 1 & & \ 1 & & \
0 & \cr \noalign{\hrule} & & \ 1 & & \ 0 & & \ 0 & & \ 1 & & \ 0 & & \
1 & & \ 0 & \cr \noalign{\hrule} & & \ 1 & & \ 0 & & \ 1 & & \ 0 & & \
0 & & \ 0 & & \ 1 & \cr \noalign{\hrule} & & \ 1 & & \ 1 & & \ 0 & & \
0 & & \ 1 & & \ 1 & & \ 0 & \cr \noalign{\hrule} & & \
1 & & \ 1 & & \ 1 & & \ 0 & & \ 0 & & \ 0 & & \ 1 & \cr
\noalign{\hrule} }}$$

\eject \vfill

\centerline{Figure Captions}

\noindent \item{} Figure~1. Some steps in the construction
of the Cantor set, with $\displaystyle h={1-\alpha\over 2}$.

\

\noindent \item{} Figure~2a. Spread of errors for rule 30. Starting
from two states $ {\bf S} $, $ {\bf S'} $ which differ in the $499$,
$500$ and $ 501$-th bits. A cross is plotted when the bits are
different. Time goes from top to bottom for $ 100 $ iterations.

\

\noindent \item{} Figure~2b. Spread of errors for rule 73. Starting
from two states $ {\bf S} $, $ {\bf S'} $ which differ in the $499$,
$500$ and $ 501$-th bits. A cross is plotted when the bits are
different. Time goes from top to bottom for $ 100 $ iterations.

\

\noindent \item{} Figure~2c. Spread of errors for rule 90. Starting
from two states $ {\bf S} $, $ {\bf S'} $ which differ in the $499$,
$500$ and $ 501$-th bits. A cross is plotted when the bits are
different. Time goes from top to bottom for $ 100 $ iterations.

\

\noindent \item{} Figure~2d. Spread of errors for rule 167. Starting
from two states $ {\bf S} $, $ {\bf S'} $ which differ in the $499$,
$500$ and $ 501$-th bits. A cross is plotted when the bits are
different. Time goes from top to bottom for $ 100 $ iterations.

\

\noindent \item{} Figure~3a. Evolution of the speed of error-profile $
{\Lambda_n \over n} $ in function of time for rule 30 showing for $
n >> 1 $ a Lyapunov exponent $ \lambda = 1 $.

\

\noindent \item{} Figure~3b. Evolution of the speed of error-profile $
{\Lambda_n \over n} $ in function of time for rule 73 showing for $
n >> 1 $ a Lyapunov exponent $ \lambda = 0 $.

\

\noindent \item{} Figure~3c. Evolution of the speed of error-profile $
{\Lambda_n \over n} $ in function of time for rule 90 showing for $
n >> 1 $ a Lyapunov exponent $ \lambda = 1 $.

\

\noindent \item{} Figure~3d. Evolution of the speed of error-profile $
{\Lambda_n \over n} $ in function of time for rule 167 showing for $
n >> 1 $ a Lyapunov exponent $ \lambda = 1 $.

\

\noindent \item{} Figure~4a. Evolution of the speed of damage
spreading $ {d_H \left( {\bf S(n)}, {\bf S'(n)} \right) \over n} $ in
function of time for rule 30. There is not a clear behavior for $ n
>> 1 $ which help to make an evaluation of the damage spreading.

\

\noindent \item{} Figure~4b. Evolution of the speed of damage
spreading $ {d_H \left( {\bf S(n)}, {\bf S'(n)} \right) \over n} $ in
function of time for rule 73. There is not a clear behavior for $ n
>> 1 $ which help to make an evaluation of the damage spreading.

\

\noindent \item{} Figure~4c. Evolution of the speed of damage
spreading $ {d_H \left( {\bf S(n)}, {\bf S'(n)} \right) \over n} $ in
function of time for rule 90. There is not a clear behavior for $ n
>> 1 $ which help to make an evaluation of the damage spreading and
is tending to zero for a complex rule.

\

\noindent \item{} Figure~4d. Evolution of the speed of damage
spreading $ {d_H \left( {\bf S(n)}, {\bf S'(n)} \right) \over n} $ in
function of time for rule 167. There is not a clear behavior for $ n
>> 1 $ which help to make an evaluation of the damage spreading and
is tending to zero for a rule which has a positive Lyapunov exponent
in our scheme. 

\

\eject \vfill

\midinsert
\vbox to 3truein{\includegraphics{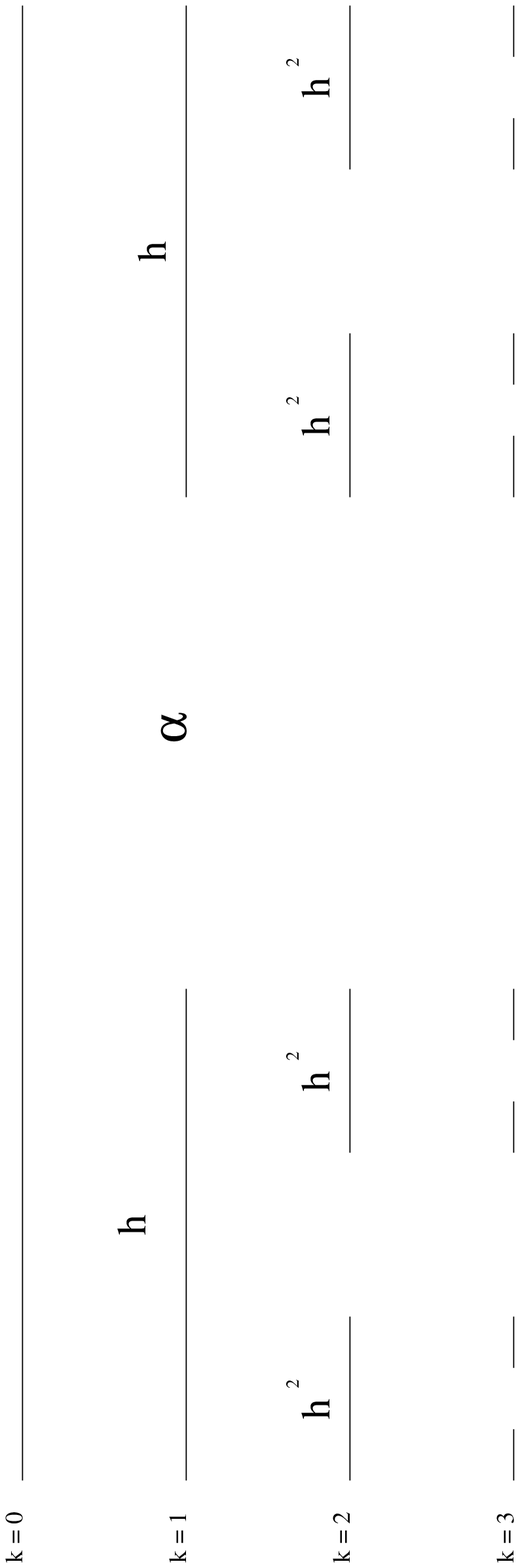}}

\noindent \item{} Figure~1. 
\endinsert

\eject \vfill

\midinsert
\vbox to 3truein{\includegraphics{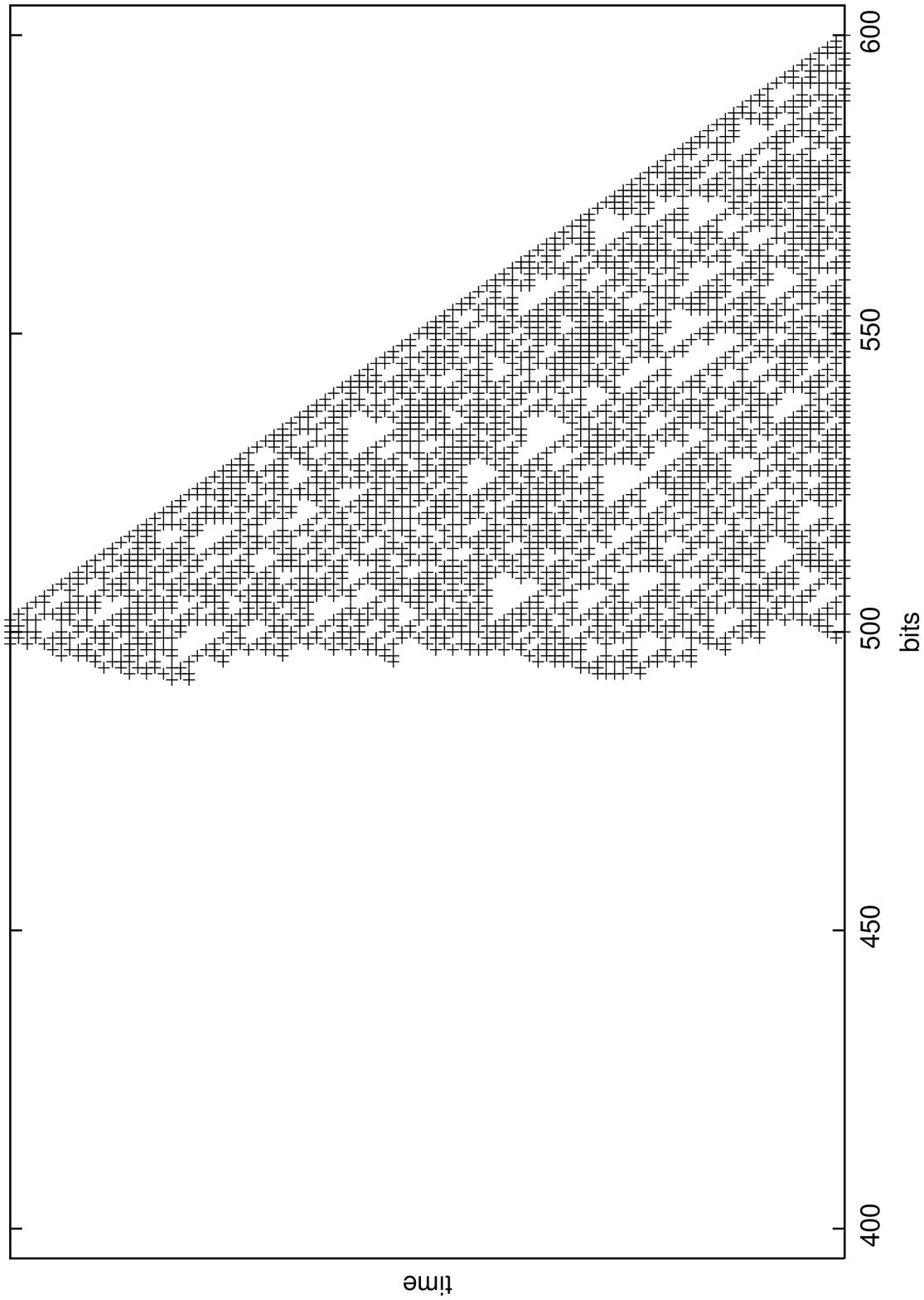}}
% \vskip 0.1cm
\noindent \item{} Figure~2a. 
\endinsert

\eject \vfill

\midinsert
\vbox to 3truein{\includegraphics{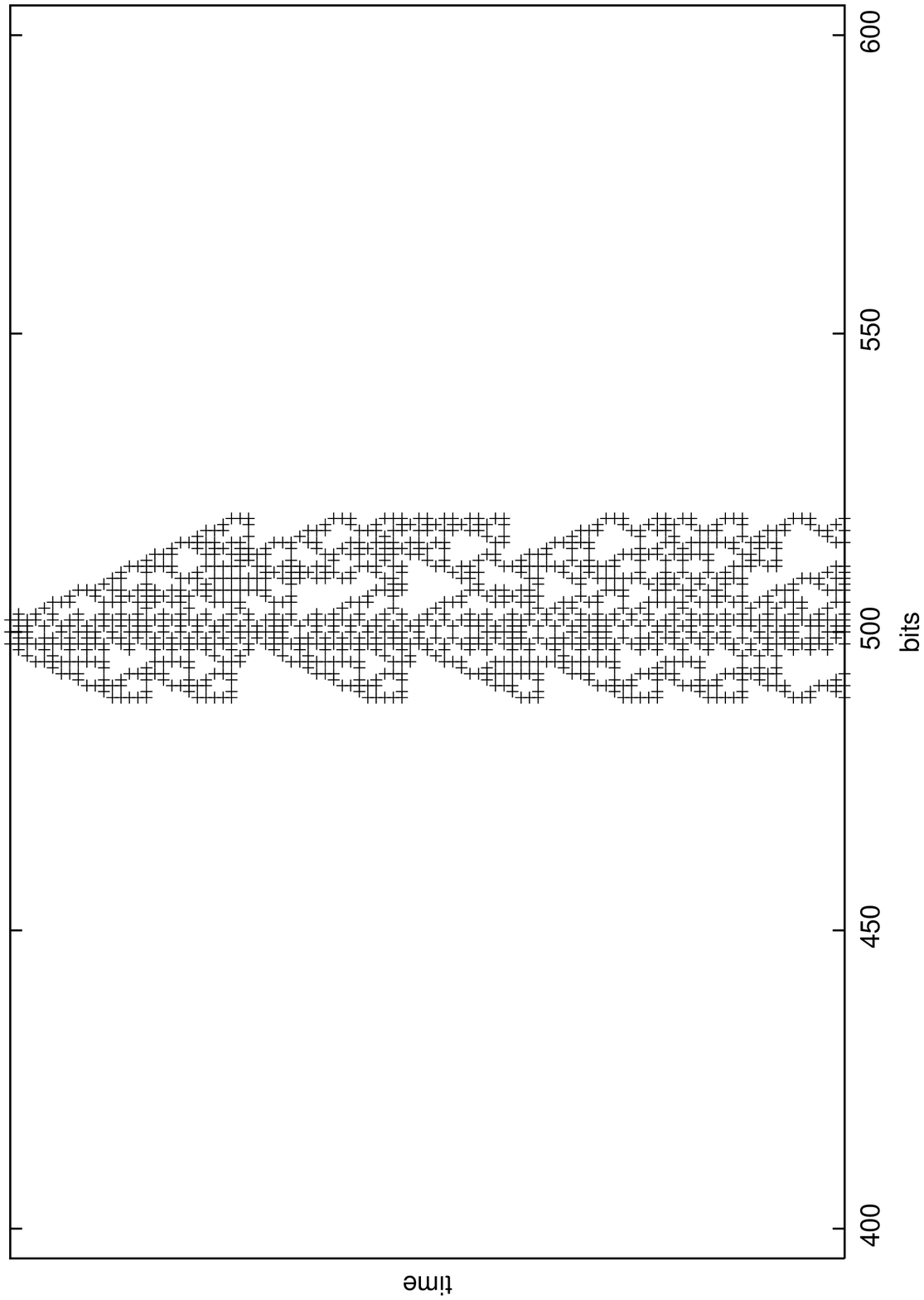}}
% \vskip 0.1cm
\noindent \item{} Figure~2b. 
\endinsert

\eject \vfill

\midinsert
\vbox to 3truein{\includegraphics{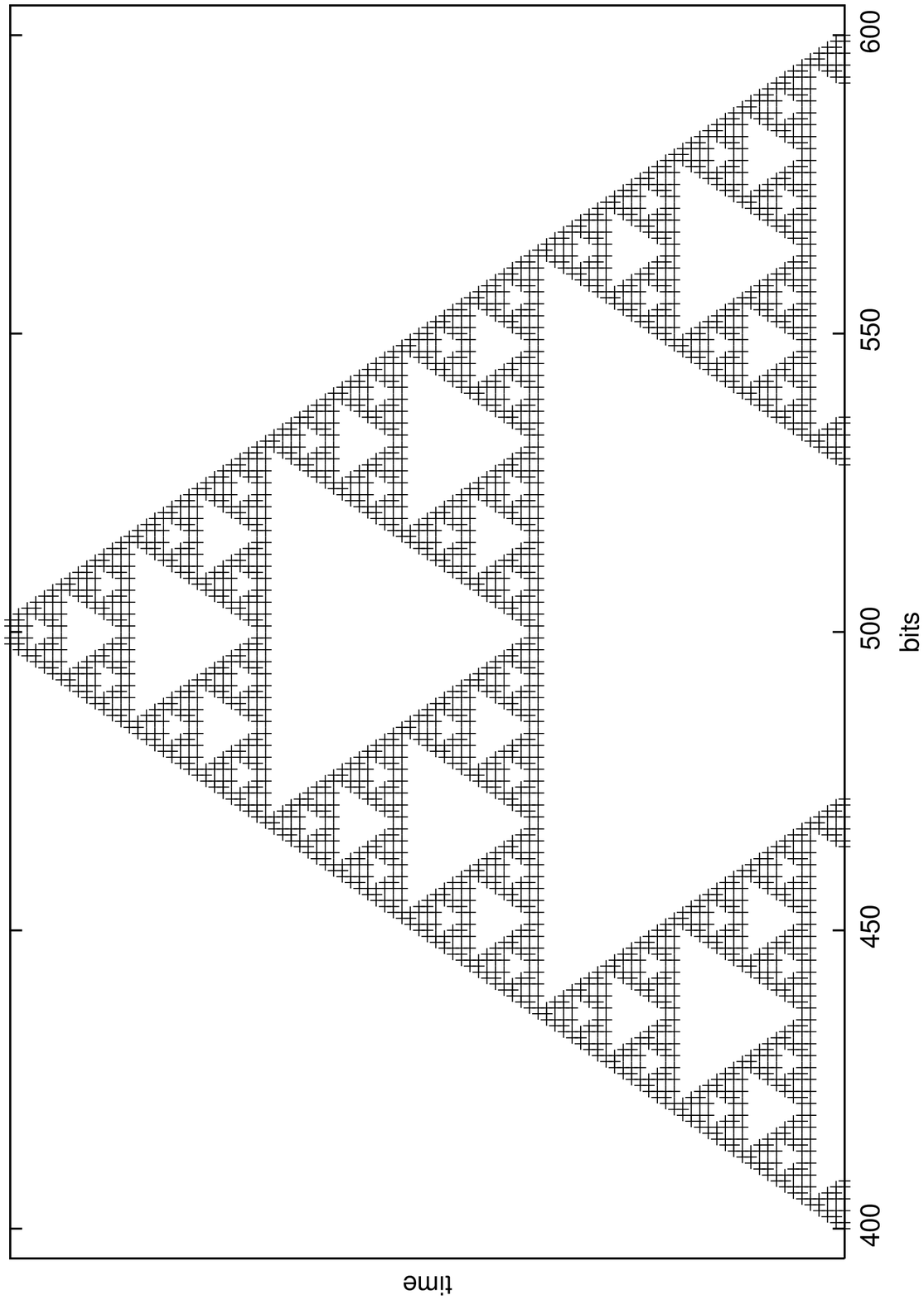}}
% \vskip 0.1cm
\noindent \item{} Figure~2c. 
\endinsert

\eject \vfill

\midinsert
\vbox to 3truein{\includegraphics{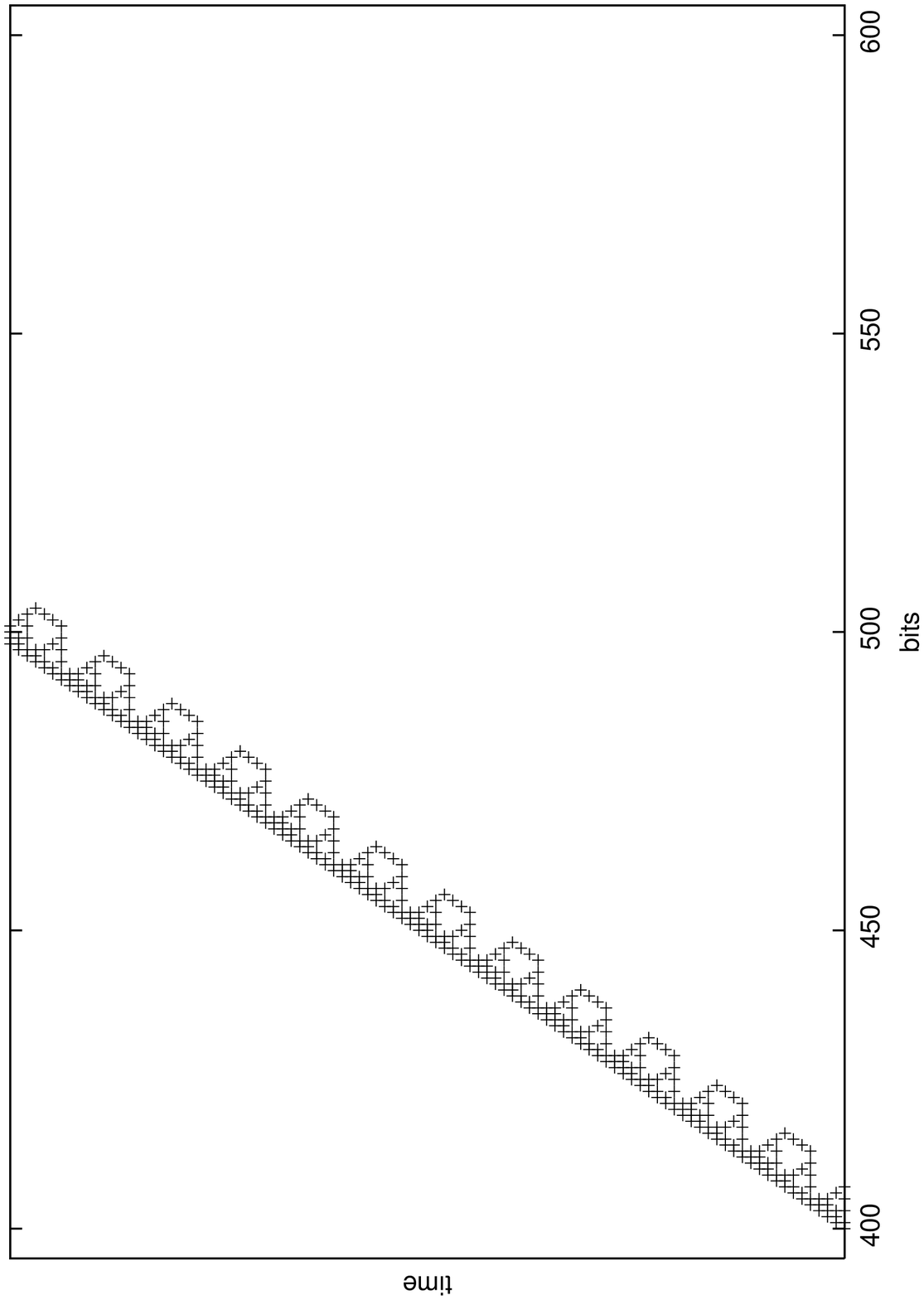}}
% \vskip 0.1cm
\noindent \item{} Figure~2d. 
\endinsert

\eject \vfill

\midinsert
\vbox to 3truein{\includegraphics{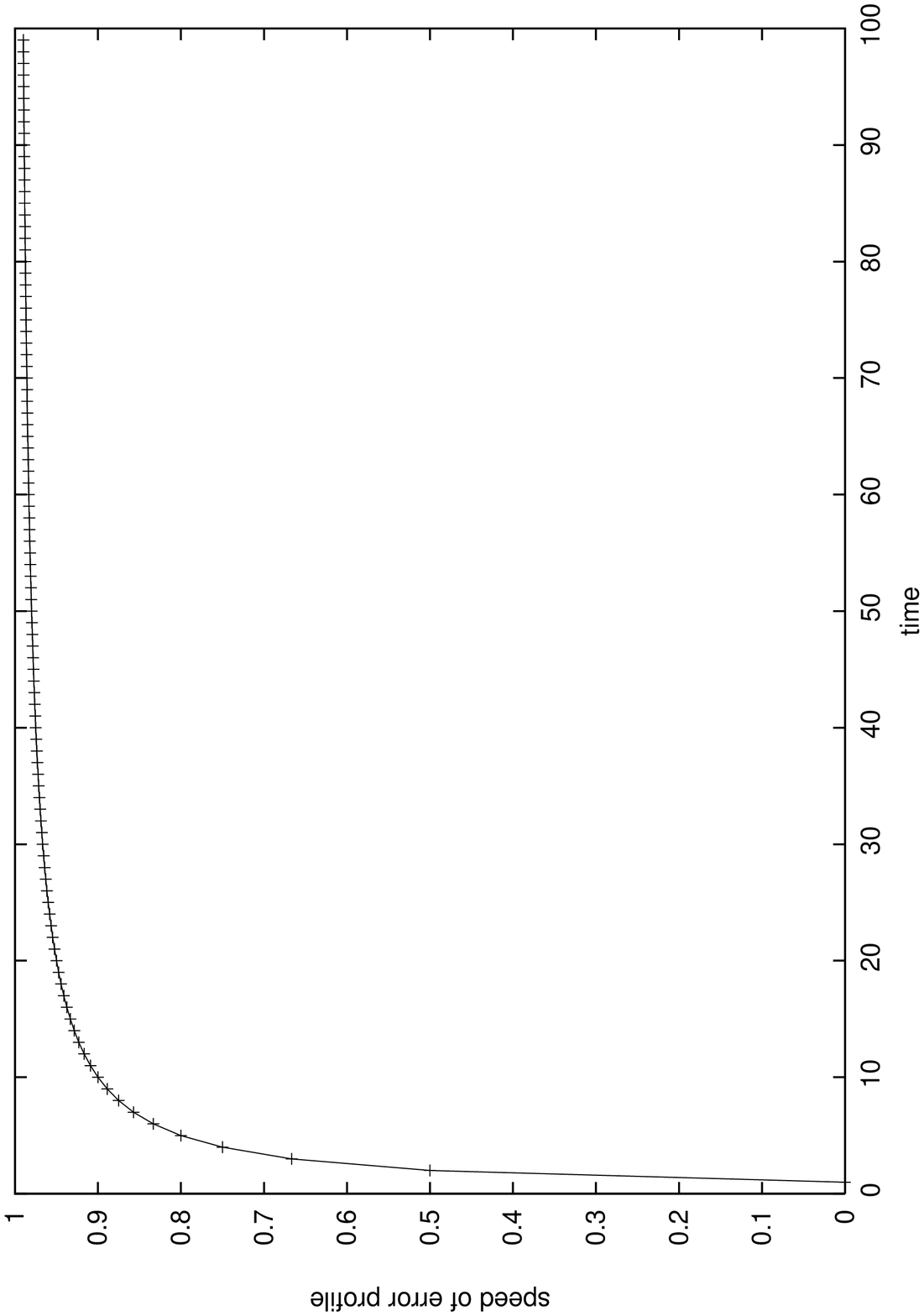}}
% \vskip 0.1cm
\noindent \item{} Figure~3a. 
\endinsert

\eject \vfill

\midinsert
\vbox to 3truein{\includegraphics{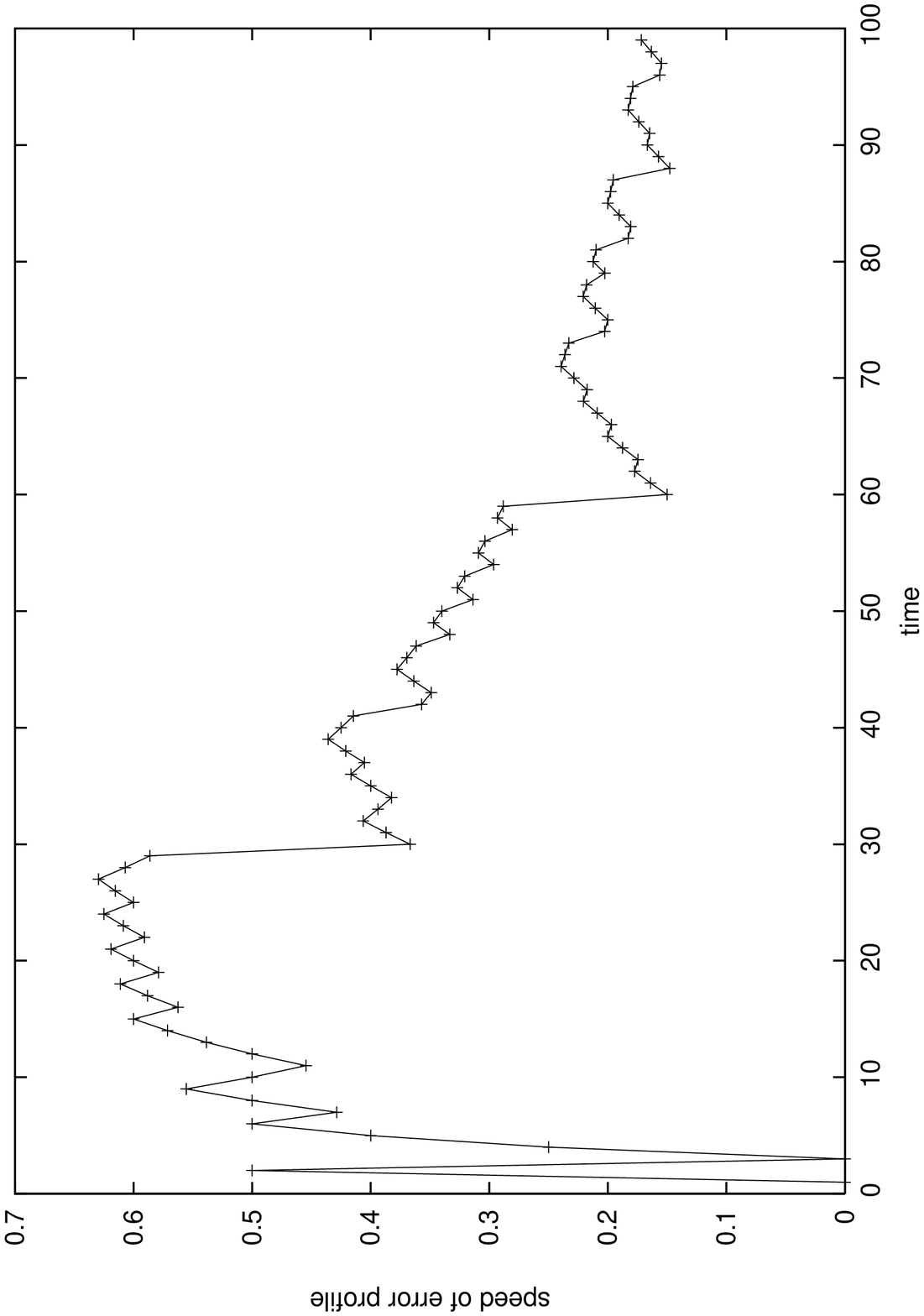}}
% \vskip 0.1cm
\noindent \item{} Figure~3b. 
\endinsert

\eject \vfill

\midinsert
\vbox to 3truein{\includegraphics{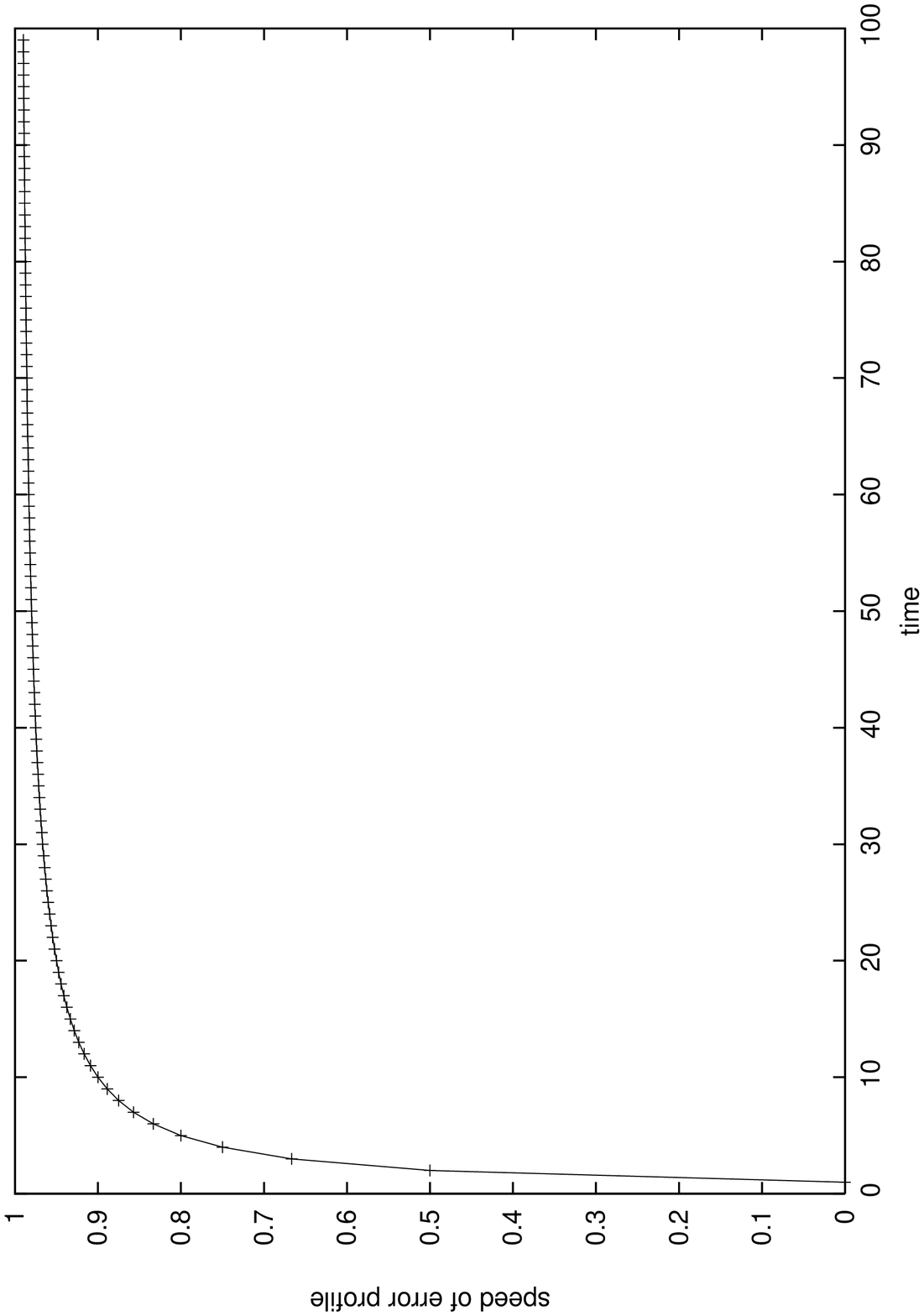}}
% \vskip 0.1cm
\noindent \item{} Figure~3c. 
\endinsert

\eject \vfill

\midinsert
\vbox to 3truein{\includegraphics{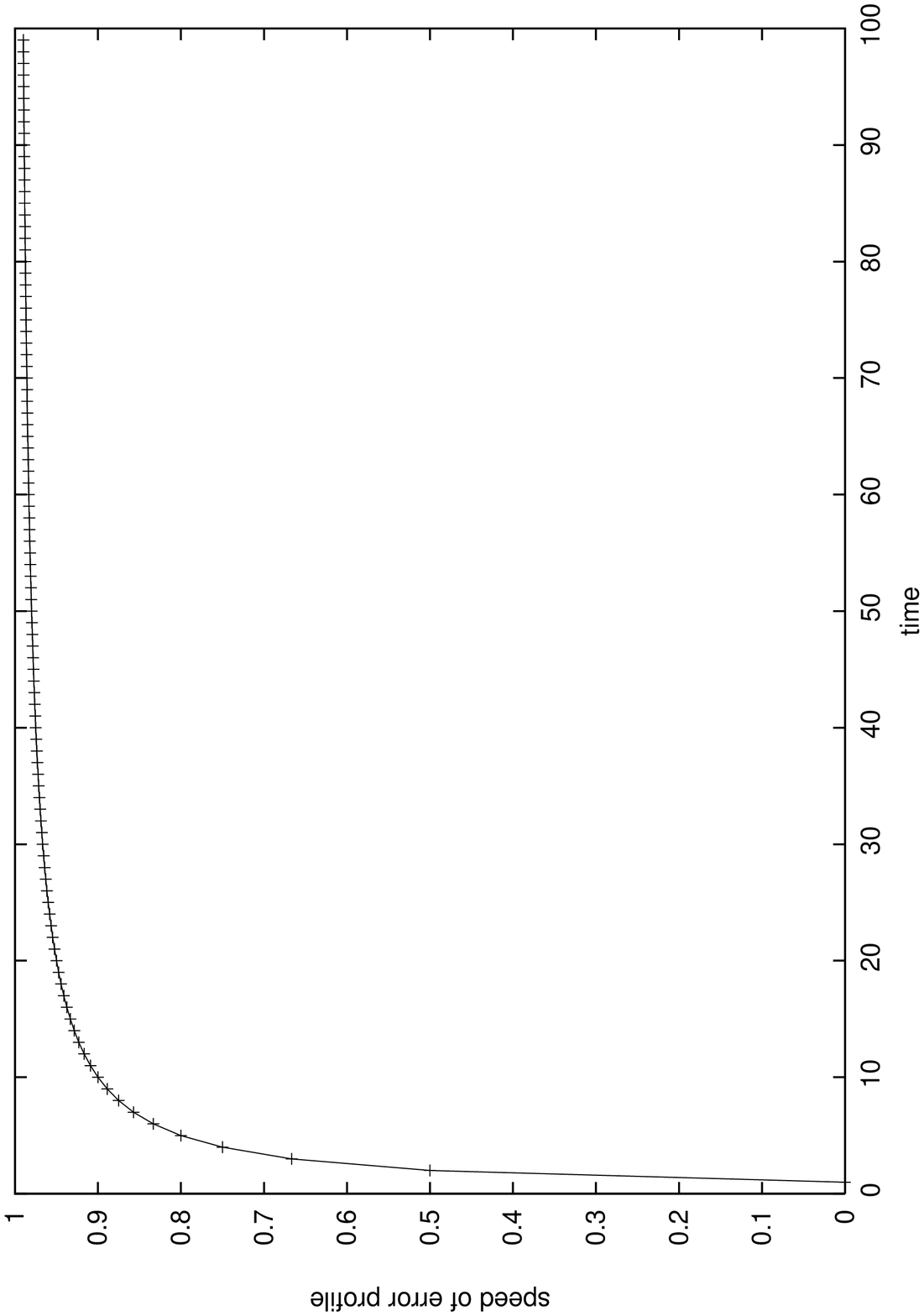}}
% \vskip 0.1cm
\noindent \item{} Figure~3d. 
\endinsert

\eject \vfill

\midinsert
\vbox to 3truein{\includegraphics{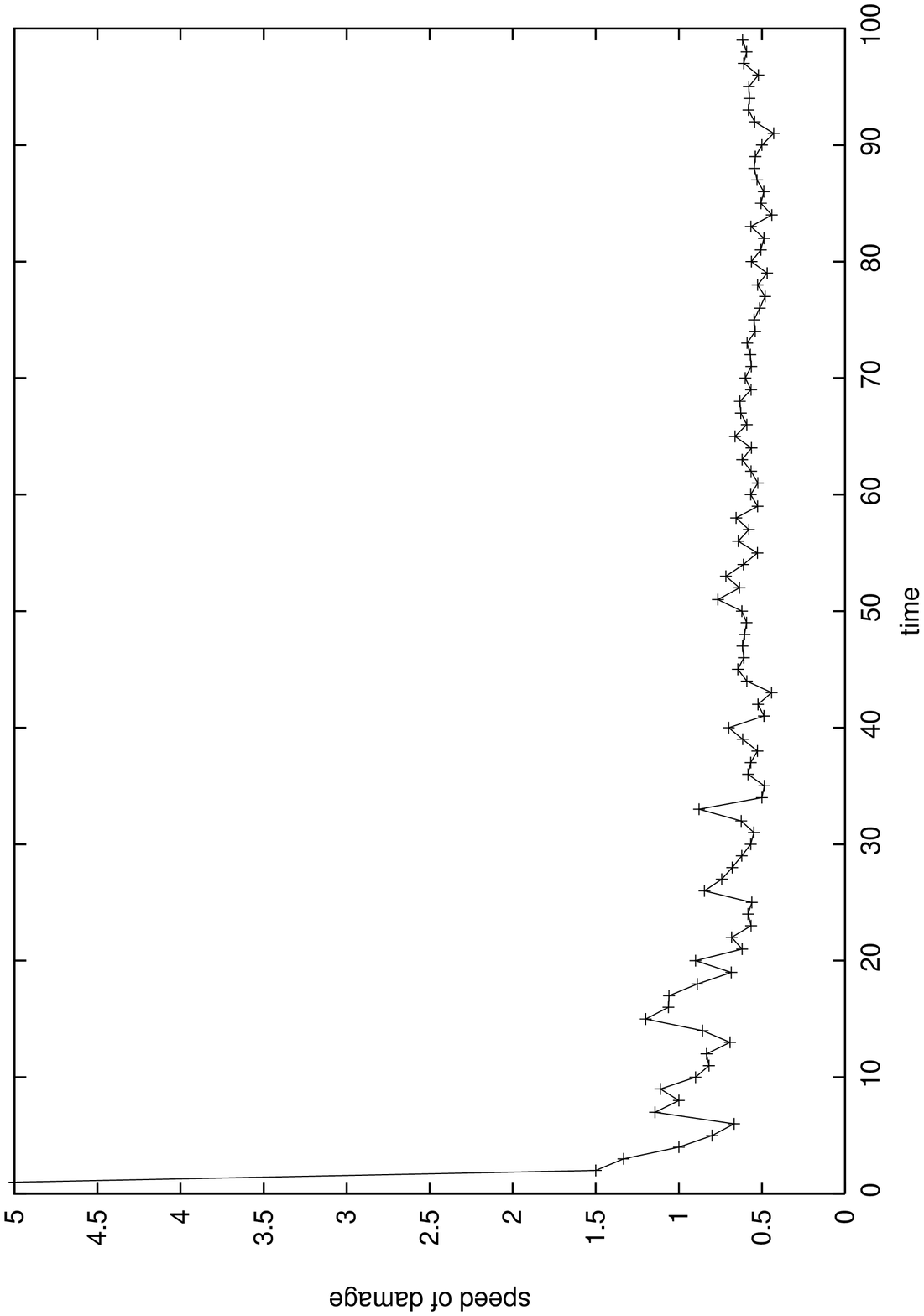}}
% \vskip 0.1cm
\noindent \item{} Figure~4a. 
\endinsert

\eject \vfill

\midinsert
\vbox to 3truein{\includegraphics{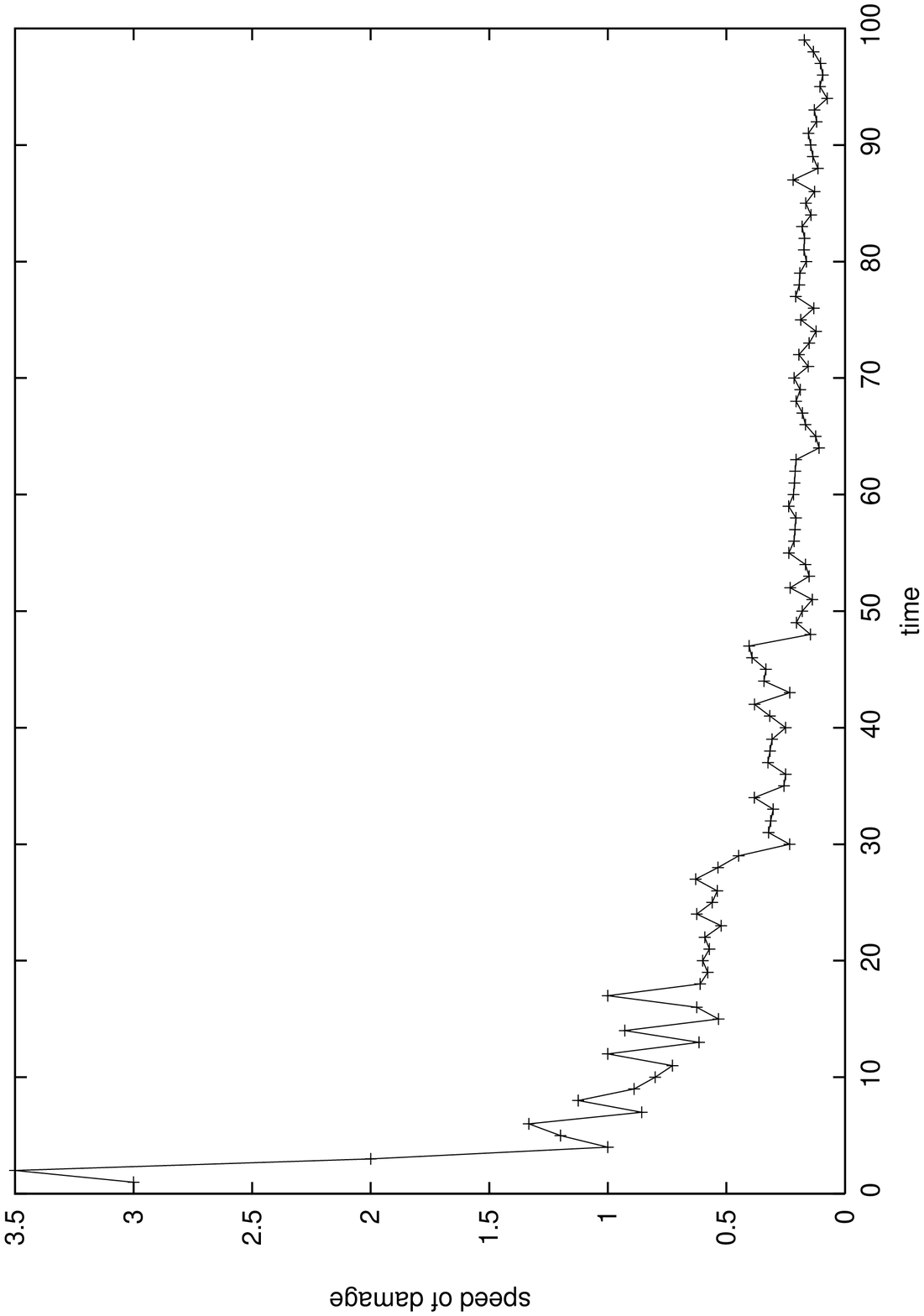}}
% \vskip 0.1cm
\noindent \item{} Figure~4b.
\endinsert

\eject \vfill

\midinsert
\vbox to 3truein{\includegraphics{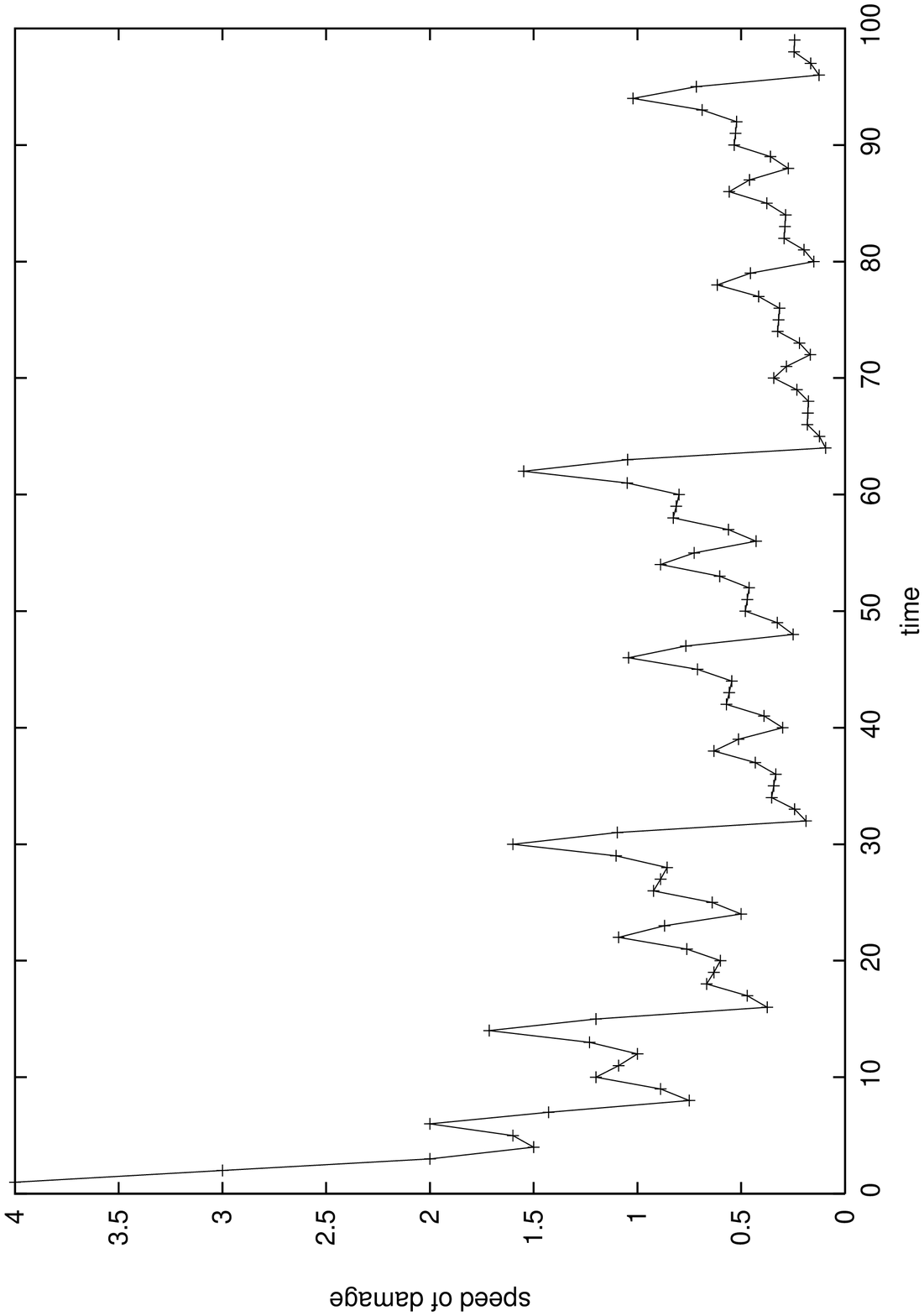}}
% \vskip 0.1cm
\noindent \item{} Figure~4c.
\endinsert

\eject \vfill

\midinsert
\vbox to 3truein{\includegraphics{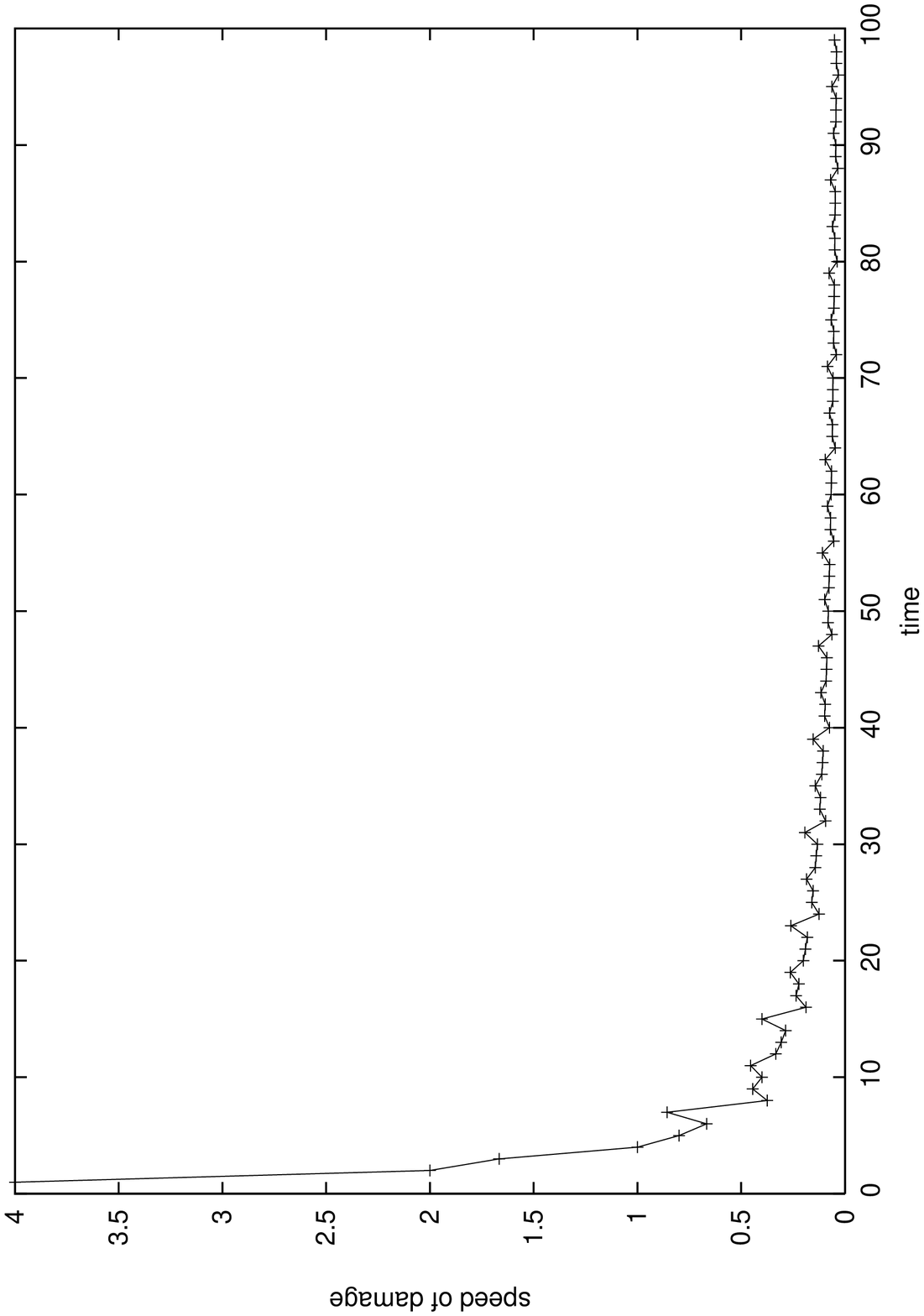}}
% \vskip 0.1cm
\noindent \item{} Figure~4d.
\endinsert

\end